\documentclass[a4paper,12pt,twoside,english]{article}
\usepackage[latin1]{inputenc}
\usepackage{parskip}               
\usepackage[dvips]{graphicx,epsfig,color}
\usepackage{wrapfig,rotating}
\usepackage{array}
\usepackage{hyperref}
\usepackage{amsfonts}       
\usepackage{amsmath}        
\usepackage{amssymb}        
\usepackage{hhline}
\usepackage{feynmp}
\usepackage{multirow}
\usepackage{cite}                  
\usepackage[ , ,bf,it]{caption}    
 
\newlength{\dinwidth}
\newlength{\dinmargin}
\newcommand{\SU}[1]{{SU(#1)}}

\newcommand{\pb}{\,\mathrm{\,pb}}
\newcommand{\fb}{\,\mathrm{\,fb}}
\newcommand{\ifb}{\,\mathrm{\,fb}^{-1}}

\newcommand{\MeV}{\,\mathrm{\,MeV}}
\newcommand{\GeV}{\,\mathrm{\,GeV}}
\newcommand{\TeV}{\,\mathrm{\,TeV}}

\newcommand{\cm}{\,\mathrm{\,cm}}
\newcommand{\mm}{\,\mathrm{\,mm}}
\newcommand{\mum}{\,\mathrm{\,\mu m}}

\newcommand{\Higgs}{H}

\newcommand{\susy}[1]{\widetilde{#1}}
\newcommand{\neutralino}[1]{\widetilde{\chi}^{0}_{#1}}
\newcommand{\chargino}[1]{\widetilde{\chi}^{\pm}_{#1}}

\newcommand{\Lepton}{L}
\newcommand{\lepton}{l}
\newcommand{\neutrino}{\nu}
\newcommand{\electron}{e}
\newcommand{\muon}{\mu}
\newcommand{\tauL}{\tau}

\newcommand{\quark}{q}
\newcommand{\uQ}{u}
\newcommand{\dQ}{d}

\newcommand{\W}{W}
\newcommand{\Z}{Z}

\newcommand{\intL}{\int \mathcal{L} \mathrm{d}t}

\newcommand{\BR}{\operatorname{BR}}

\newcommand{\Rslash}   {\ensuremath{ R\kern-0.5em\slash   }}

\begin{document}
\begin{titlepage}
  \begin{flushleft}
    {\tt DESY 13-128    \hfill    ISSN 0418-9833} \\
    {\tt LC-PHSM-2013-024}                \\
    {\tt December 2013}                  \\
  \end{flushleft}

  \vspace{1.0cm}
  \begin{center}
    \begin{Large}
      {\bfseries \boldmath Bilinear $R$ Parity Violation at the ILC -- Neutrino Physics at Colliders}

      \vspace{1.5cm}
      Benedikt Vormwald$^{1,2}$ and Jenny List$^1$
    \end{Large}

    \vspace{.3cm}
    1- Deutsches Elektronen-Synchrotron DESY \\ 
    Notkestr. 85,  22607 Hamburg, Germany
    \vspace{.1cm} \\
    2- Universit\"at Hamburg, Institut f\"ur Experimentalphysik \\
    Luruper Chaussee 149,  22761 Hamburg, Germany
  \end{center}

  \vspace{1cm}

  \begin{abstract}
Supersymmetry (SUSY) with bilinearly broken $R$ parity (bRPV) offers an attractive possibility to explain the origin of neutrino masses and mixings.
In such scenarios, the study of neutralino decays at colliders gives access to neutrino sector parameters.

The ILC offers a very clean environment to study the neutralino properties as well as its subsequent decays, which typically involve a $W$ or $Z$ boson and a lepton.
This study is based on ILC beam parameters according to the Technical Design Report for a center of mass energy of $500\,\mathrm{GeV}$.
A full detector simulation of the International Large Detector (ILD) has been performed for all Standard Model backgrounds and for neutralino pair production within a simplified model.
The bRPV parameters are fixed according to current neutrino data.

In this scenario, the $\tilde{\chi}^0_1$ mass can be reconstructed with an uncertainty of $\delta m^{\mathrm{fit}}_{\neutralino{1}}=(40\text{(stat.)} \oplus 50\text{(syst.)})\MeV$ for an integrated luminosity of $500\,\mathrm{fb}^{-1}$ from direct $\tilde{\chi}^0_1$ pair production, thus, to a large extent independently of the rest of the SUSY spectrum.
The achievable precision on the atmospheric neutrino mixing angle $\sin^2 \theta_{23}$ from measuring the neutralino branching fractions BR($\tilde{\chi}_1^0\rightarrow W \mu$) and
BR($\tilde{\chi}_1^0\rightarrow W \tau$) at the ILC is in the same range than current uncertainties from neutrino experiments.
Thus, the ILC could have the opportunity to unveil the mechanism of neutrino mass generation.

  \end{abstract}

\end{titlepage}

\section{Introduction}
Supersymmetry (SUSY) \cite{Wess:1973kz,Wess:1974tw} is a very appealing extension of the Standard Model (SM).
It provides an elegant solution for the Higgs hierarchy problem, makes gauge 
unification possible and apart from that, SUSY is the only non-trivial extension of the Lorentz algebra \cite{Haag:1974qh}.
In the most general renormalizable Lagrangian of the Minimal Supersymmetric Standard Model (MSSM) trilinear and bilinear terms appear, which violate the conservation of
baryon number $B$ and lepton number $L$. 

The presence of all these terms would lead to proton decay, which is experimentally not observed.
A common way to circumvent this problem is to introduce a discrete $\mathbb{Z}_2$ symmetry 
assigned to each field in order to suppress these terms. This quantum number, called $R$ parity, has the form
\begin{eqnarray}
R = (-1)^{3B+L+2S},
\end{eqnarray}
where $B$ is the baryon number, $L$ the lepton number and $S$ the spin of the field. Hence, SM particles always carry $R=+1$ and SUSY particles $R=-1$.
The conservation of this quantum number has the consequence, that all $B$ and $L$ breaking
terms in the SUSY Lagrangian are forbidden and the proton remains stable.

However, proton decay only appears if $B$ and $L$ violation is present at the same time.
So, breaking either $B$ or $L$ is well consistent with proton stability.
Thus, $R$ parity violating (RPV) SUSY scenarios are also viable alternatives to the widely studied $R$ parity conserving (RPC) scenarios.

We will focus in the following on bilinear $R$ parity violation (bRPV), which has the interesting
feature to be able to introduce neutrino masses and mixings. The phenomenology of this mechanism has already been
discussed in detail in the literature \cite{Romao:1991ex, Mukhopadhyaya:1998xj, Choi:1999tq, Romao:2000hv, Porod:2000hv, Hirsch:2003fe}. Our aims is to investigate the performance of the International Linear Collider and one of its proposed detector concepts for measuring the atmospheric neutrino mixing angle in such a bRPV SUSY scenario, based on full detector simulation and current beam parameters~\cite{Behnke:2013lya, Adolphsen:2013kya}.

In this section, we briefly summarise the basic concept of bilinear $R$ parity violation and its connection to collider physics.
The superpotential and the corresponding soft SUSY breaking terms in bRPV SUSY have the form
\begin{eqnarray}
W^{\mathrm{MSSM}} =& W^{\mathrm{MSSM}}_{\mathrm{RPC}} + \epsilon_i \Lepton_i \Higgs_\uQ, \\
\mathcal{L}^{\mathrm{MSSM}}_{\mathrm{soft}} =& \mathcal{L}^{\mathrm{MSSM}}_{\mathrm{soft},{RPC}} + \epsilon_i B_i L_i H_u,
\end{eqnarray}
where $i=\{\electron,\muon,\tauL\}$ is the generation index. $H_u$ indicates the $\SU{2}$ doublet of the Higgs superfield and $L_i$ the $\SU{2}$ doublet of the lepton superfield. 
$\epsilon_i$ and $B_i$ are bRPV parameters.
In addition to that, the three sneutrinos acquire a vacuum expectation value (VEV) $\langle\susy{\neutrino}_i\rangle=v_i$.
Because of three additional tadpole equations one ends up with six free parameters for bilinear $R$ parity violation. These parameters can be fixed by fitting them to neutrino observables,
like neutrino mass differences and mixing angles.

The introduction of lepton number violation allows the neutrinos to mix with the other neutral
fermions of the model, i.e.~the gauginos and higgsinos. Thus, in
the basis of neutral fermions $\Psi^{0T} =
\left(\susy{B},\susy{W}^0,\susy{H}^0_\dQ,\susy{H}^0_\uQ,\neutrino_e,
\neutrino_\mu , \neutrino_\tau\right)$ the corresponding
mass term in the Lagrangian looks like
\begin{eqnarray}
\mathcal{L}=-\frac{1}{2}\left(\Psi^0\right)^T \mathbf{M}_N \Psi^0 + c.c.,
\end{eqnarray}
where the mass matrix $\mathbf{M}_N$ has additional off--diagonal entries due to bilinear
$R$ parity breaking.

Diagonalising $\mathbf{M}_N$ generates one neutrino mass at tree
level as well as two neutrino mixing angles. 
The atmospheric neutrino mixing angle, for instance, writes as
\begin{eqnarray}
\tan(\theta_{23})&=&\frac{\Lambda_\mu}{\Lambda_{\tau}}, \label{eq:theta23}
\end{eqnarray}
where $\Lambda_i = \mu v_i + v_d \epsilon_i$ are so called alignment parameters.
Herein, $\mu$ is the MSSM higgsino mass parameter and $v_d$ represents the VEV of the down-type Higgs.
It has been show in \cite{Romao:2000hv} that the remaining neutrino mixing angle and neutrino masses can be derived on 1-loop level.

A very interesting feature of this model is that the 
left-handed part of the $\neutralino{1}-\W-\lepton_i$--coupling is approximately 
proportional to the alignment parameters
\begin{eqnarray}
O_{\neutralino{1}Wl_i}\simeq\Lambda_i\cdot f(M_1, M_2, \mu, v_d, v_u)\propto\Lambda_i, \label{eq:coupling}
\end{eqnarray}
where $f$ is a function of the soft SUSY breaking parameters. The full expression of $O_{\neutralino{1}Wl_i}$ can be found in \cite{Porod:2000hv}.
Combining eq.~\eqref{eq:theta23} with eq.~\eqref{eq:coupling} makes
clear that neutrino mixing can be determined from measuring branching rations of
the neutralino decays:
\begin{eqnarray}
\tan^2(\theta_{23}) \simeq \frac{O_{\neutralino{1}W\muon}^2}{O_{\neutralino{1}W\tauL}^2} = \frac{\BR(\neutralino{1}\rightarrow\W\mu)}{\BR(\neutralino{1}\rightarrow\W\tau)} \label{eq:theta23fromBR}
\end{eqnarray}

The exact relation only holds at tree level. Ref. \cite{Porod:2000hv} shows that via loop contributions additional SUSY parameters enter into the mass matrix and thus into the relation between branching ratios and neutrino mixing angles. For this study, we use \texttt{SPheno3.2.4beta}~\cite{spheno324beta} to fit the alignment parameters to current neutrino data and to extract the physical observables, in particular the neutralino branching ratios, at loop level. This procedure is repeated for every considered point in SUSY parameter space. 
 
It is worth mentioning that for bRPV SUSY there is always a connection between LSP decays and neutrino physics independently of the type of the LSP, which is shown in \cite{Hirsch:2003fe}.

The remainder of the paper is organised as follows:
In section~\ref{sec:lhc} we summarise the status of RPV searches at the LHC.
In the following section~\ref{sec:ilcstudy} we focus on the study at the ILC.
Here we introduce the studied simplified scenario, give some details on the used Monte-Carlo samples, define the event selection and comment on sources for systematic uncertainties.
The results are presented in section~\ref{sec:results}, where we discuss the LSP mass measurement, the expected signal significance in the parameter space of the simplified model and the precision in measuring the atmospheric mixing angle from the ratio of two LSP branching ratios.
We finally conclude by summarising our obtained results in section~\ref{sec:conclusions}.

\section{Status at the LHC} \label{sec:lhc}
In pp collisions the dominant SUSY production mode is via squark and gluino production.
Those coloured particles then decay via cascades down to the lightest SUSY particle (LSP), which in the RPV case 
then decays into Standard Model particles. The main difference to studies assuming $R$ parity conservation is that the cut on missing energy is relaxed
significantly, since the LSP does not escape undetected anymore.

The ATLAS collaboration performed a dedicated bRPV SUSY search in the framework
of the CMSSM, where the RPV parameters have been
fitted to neutrino data and have not been taken as free \cite{ATLAS:2012dp}.
In this study contributions from all possible production modes have been taken into account.
This study excludes a wide range of the CMSSM parameter plane reaching up to $m_{1/2}\approx600\GeV$ or $m_0\approx1.2\TeV$.
However, most of the exclusions of the parameter space result from limits on 
coloured particles for the specific parameter points.

Except for this, various RPV SUSY searches have been performed in the simplified model framework
\cite{LHC:rpvsearches1, LHC:rpvsearches2, LHC:rpvsearches3, LHC:rpvsearches4, LHC:rpvsearches5, LHC:rpvsearches6, LHC:rpvsearches7}.
Many of these studies assume strong production, which is dominant for not too high squark or gluino
masses. 
So, the derived limits are again predominantly limits on the coloured sector of the model and the electroweak sector remains untested.

The LHC now starts to become sensitive to
direct electroweakino production and is able to set limits on electroweakino masses 
in the RPC \cite{LHC:directproduction1, LHC:directproduction2, LHC:directproduction3} 
as well as in the RPV case \cite{LHC:rpvdirectsearches}. However, the cross section for direct $\neutralino{1}\neutralino{1}$ production at the LHC is extremely small, unless a specific Higgsino-Wino mixing is assumed~\cite{LHC:rpvdirectsearches}. So, usually a production of heavier electroweakinos is considered, which decay via the LSP to Standard Model particles.
Therefore, the limits can only be given as a combination of $\neutralino{1}$ mass and the mass of a heavier electroweakino.
In the RPV case, currently one study is present that assumes direct chargino production and one additional non-vanishing trilinear
RPV coupling \cite{LHC:rpvdirectsearches}. The resulting limits in the $m_{\neutralino{1}}$-$m_{\chargino{1}}$ plane strongly depend on the assumed RPV coupling.
Under the most optimistic assumptions, chargino masses up to $750\GeV$ have been probed.
Due to the strong dependency on the assumed type
and strength of the RPV couplings, it is not possible to directly re-interpret these limits in a bRPV scenario which accounts for neutrino data.

%
\section{Bilinear RPV at the ILC} \label{sec:ilcstudy}
\subsection{Model definition}
At the ILC the situation is complementary to the LHC: Here, direct electroweakino production is dominant
and the electroweak sector can be probed directly, including LSP pair production.

\begin{figure}
 \centering
 \includegraphics[width=0.34\textwidth]{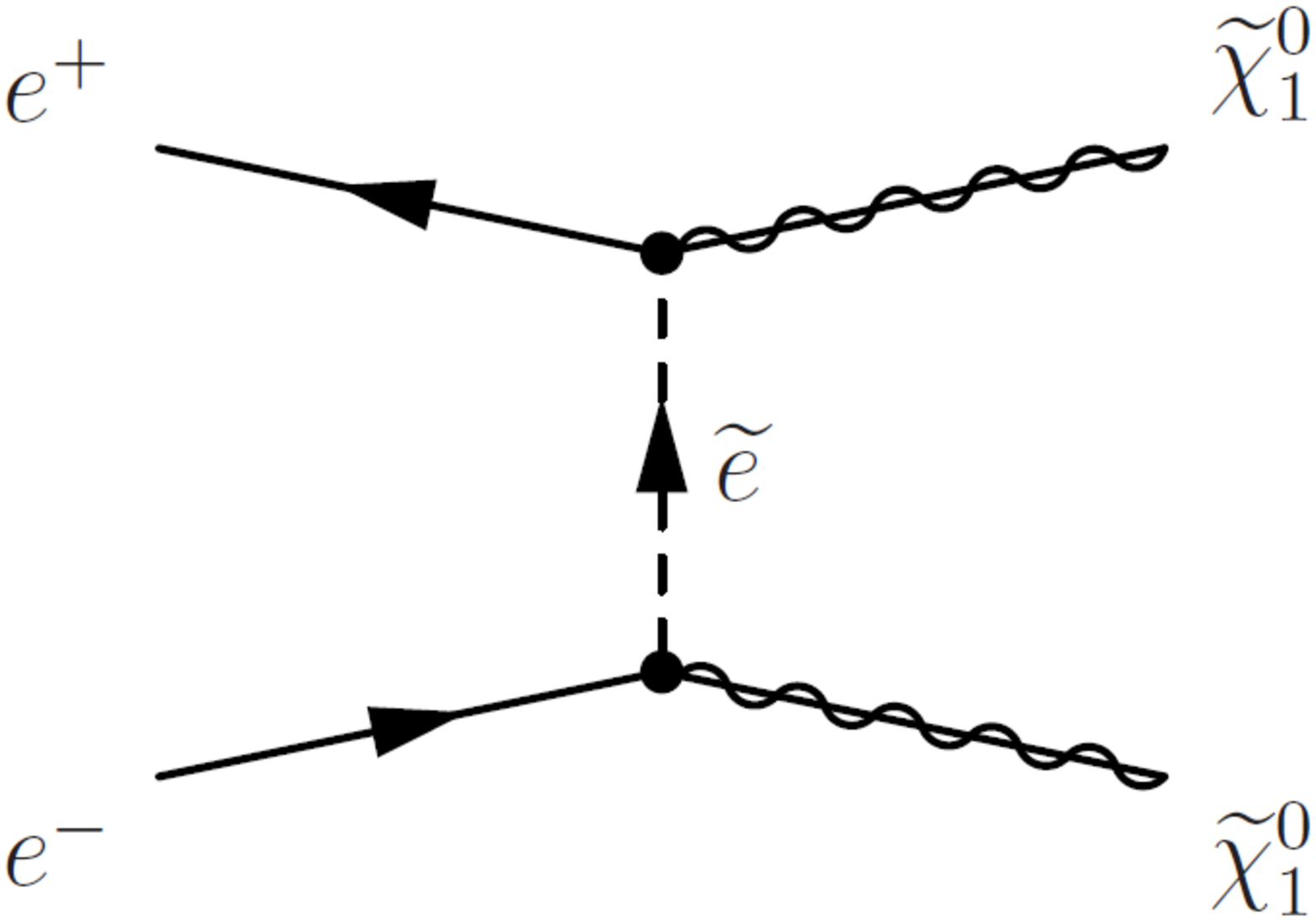}
 \hspace{1cm}
 \includegraphics[width=0.34\textwidth]{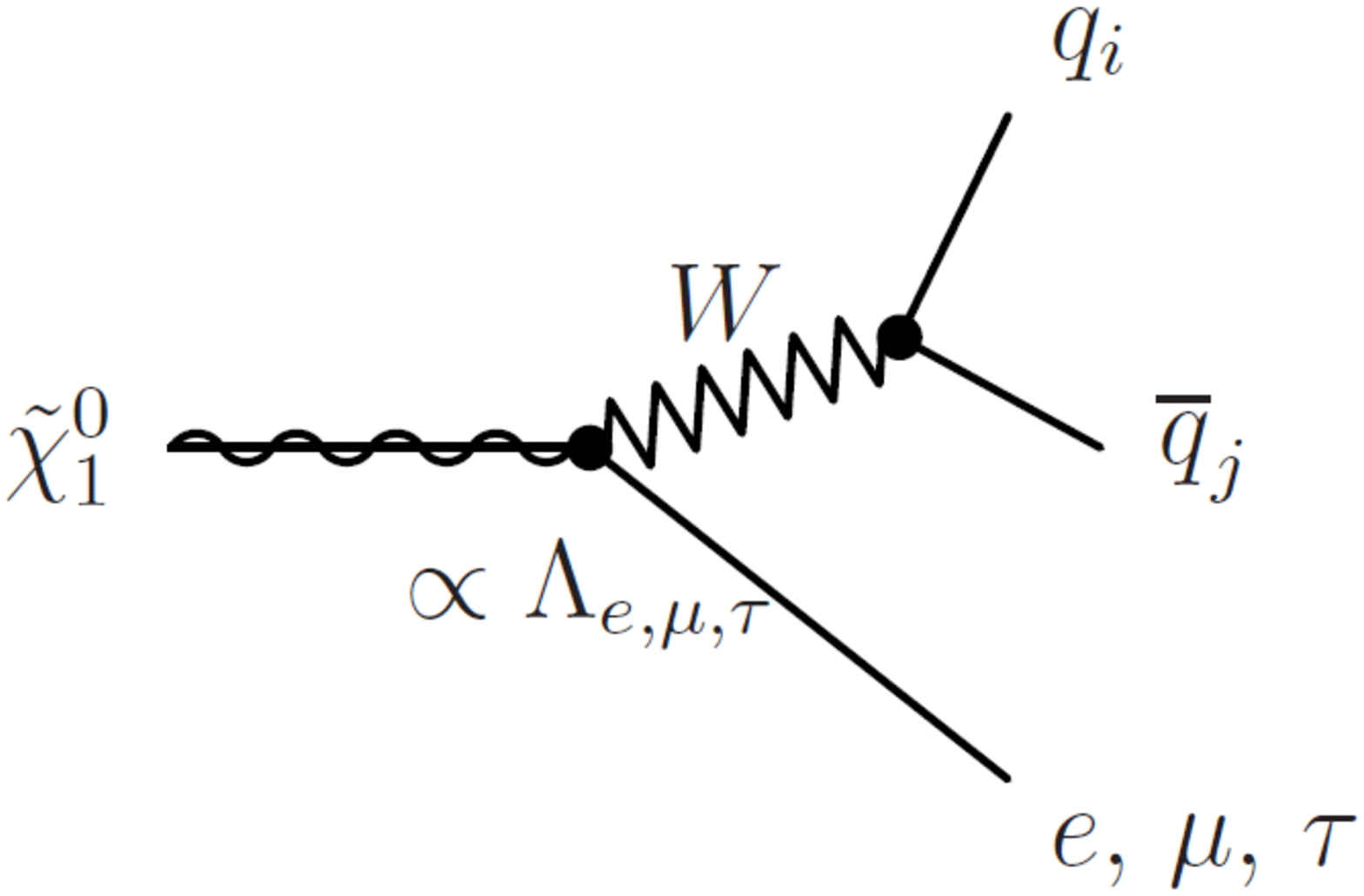}
%
%
  \caption{Left: Main production channel for an bino-like LSP at the ILC: t-channel exchange of an selectron. Right: LSP decay to an on-shell $\W$ boson and a lepton offers direct access to bRPV alignment parameters $\Lambda_i$ that account for neutrino mixing.}
  \label{fig:productiondecay}
 \end{figure}

In our scenario we assume the lightest neutralino to be a bino, which 
leaves only the $t/u$-channel production for direct $\neutralino{1}$ pair production (see fig.~\ref{fig:productiondecay}).
In presence of bRPV couplings, not only selectrons are possible as exchange particle, but
due to the additional terms in principle all other charged scalars could contribute.
However, these contributions are strongly suppressed by the small RPV couplings.
We define a simplified model in which we set all masses of the SUSY particles to the
multi-TeV scale, except for $m_{\neutralino{1}}$ and $m_{\susy{\electron}_R}$.
Thus, those two parameters fix the production cross section.

In the case of a wino LSP, the cross section would drop due to the missing coupling to the right selectron.
For a light left selectron, however, the situation is comparable to the bino case.
A higgsino-like LSP would allow $s$-channel associate production of $\neutralino{1}$ and $\neutralino{2}$, 
predominantly via a $\Z$ boson. The cross section for this production process is about $100\fb - 200\fb$ \cite{Baer:2011ec}.
In a light higgsino scenario $\neutralino{1}$ and $\neutralino{2}$ are usually close in mass and
the decay products of $\neutralino{2}$ to $\neutralino{1}$ are rather soft. 
Thus, experimentally the situation is comparable to direct $\neutralino{1}$ pair production.

\begin{figure*}[htb]
 \centering
 \includegraphics[width=0.496\textwidth]{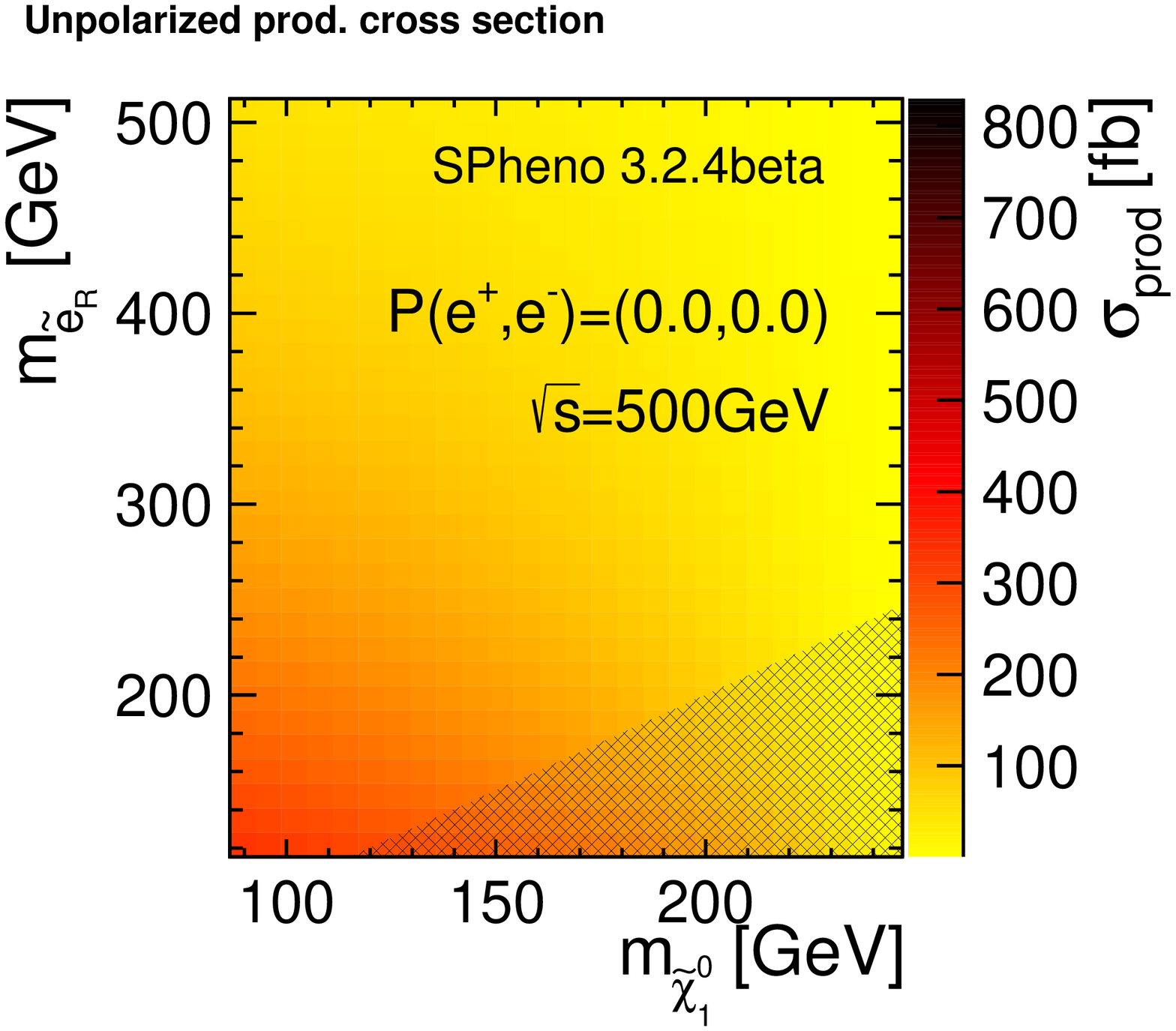}
 \includegraphics[width=0.496\textwidth]{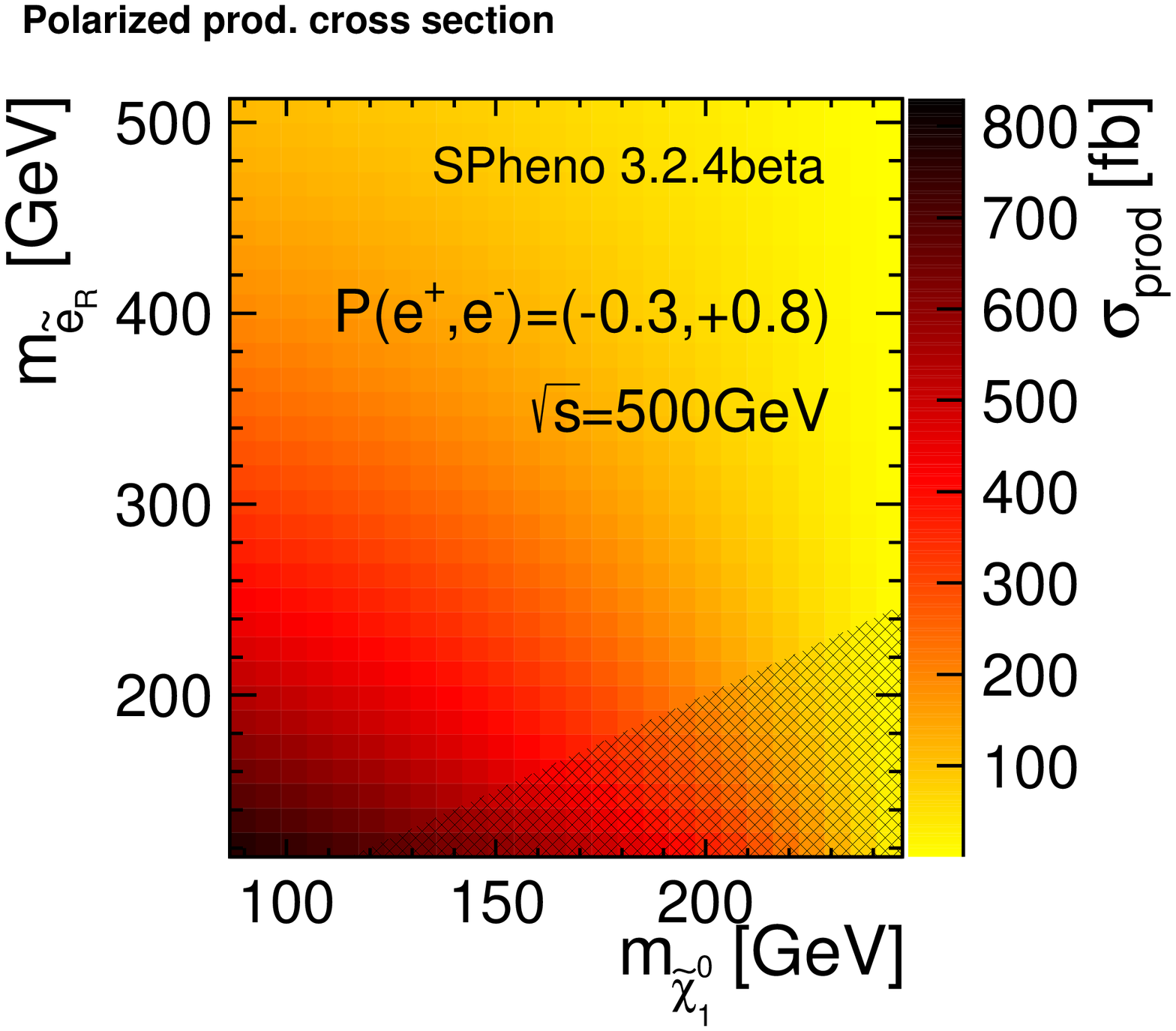}
 \caption{Unpolarised and polarised production cross section at the ILC500 in the described simplified model. 
          Beam polarisation can significantly enhance the production cross section. The shaded area shows the region of the parameter space, where the selectron becomes the LSP.}
 \label{fig:productioncrosssection}
\end{figure*}

At the ILC, the electron beam is polarised to $80\%$ and the positron beam to up to $60\%$ \cite{Adolphsen:2013kya}.
The advantages of beam polarisation have been discussed in many studies. A comprehensive overview can be found, for instance, in \cite{MoortgatPick:2005cw}.
In the case of $t/u$-channel production with selectron exchange different combinations of beam polarisation influence
the production cross section significantly. Figure~\ref{fig:productioncrosssection} shows the cross section in
the $\tilde{e}_R$-$\tilde{\chi}^0_1$ mass plane for unpolarised beams (left) and for the baseline polarisation
of $P(e^+,e^-)=(-30\%,+80\%)$ (right), which enhances the cross section considerably.

\begin{figure*}
 \centering
 \includegraphics[width=0.49\textwidth]{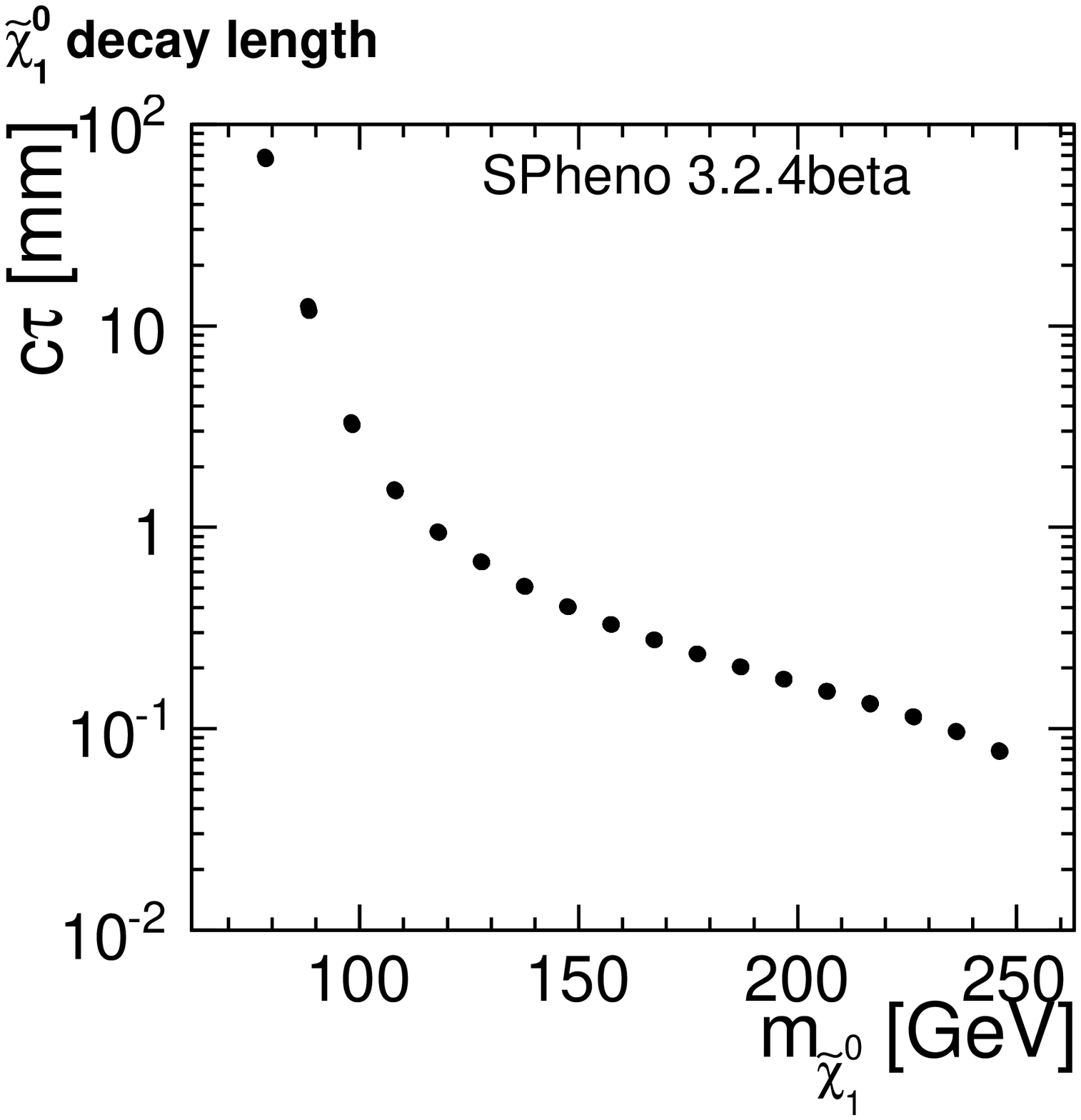}
 \includegraphics[width=0.45\textwidth]{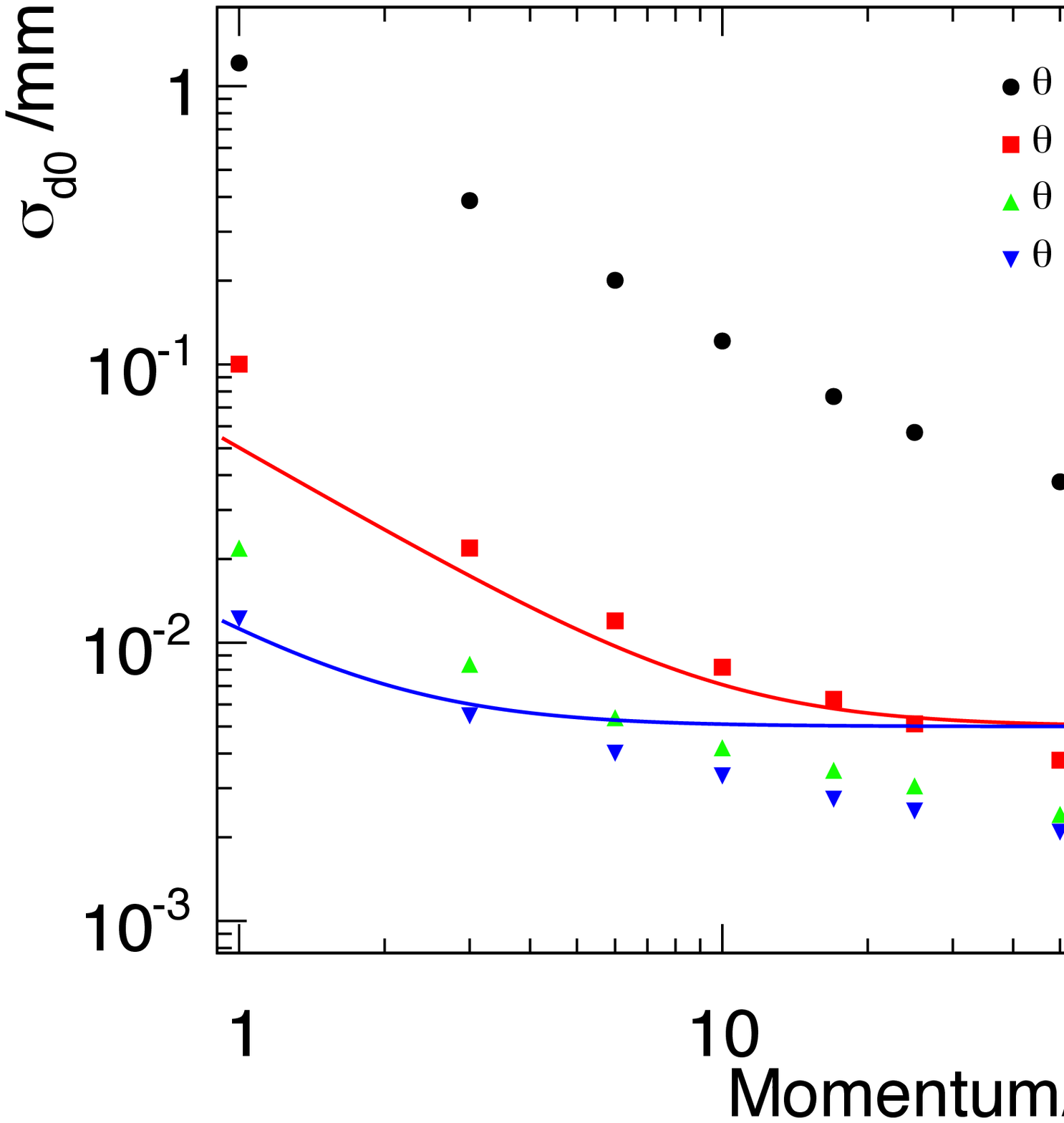}
 \caption{Left: Decay length of the LSP in dependence of its mass in the simplified model. For a light spectrum of the remaining sparticles, the decay length can be smaller.
          Right: Impact parameter resolution of the ILD detector
  concept (from~\cite{Behnke:2013lya}).}
 \label{fig:xseclength}
\end{figure*}

Since the RPV couplings are very small if they are used to describe neutrino data correctly,
light LSPs can become rather long-lived with a decay length of meters up to kilometers and escape the detector.
Therefore, scenarios with very light LSPs would behave very similar to RPC SUSY scenarios, where the LSPs are stable.
However, those escaping pair-produced LSPs could still be detected via radiative LSP production for cross sections of $\mathcal{O}(10\fb)$ as demonstrated in \cite{Bartels:2012ex}.
In this study, we focus on scenarios where the on-shell $\W$-decay channel is available, i.e.~on LSP masses larger than the $W$ mass.
Thus, the decay length of the LSP lies between $10\cm$ and $100\mum$ as depicted in fig.~\ref{fig:xseclength} (left).
The right-hand panel fo figure~\ref{fig:xseclength} shows the impact parameter resolution of the ILD detector concept as determined from full detector simulation~\cite{Behnke:2013lya}.
It indicates that the arising displaced vertices in this scenarios are well detectable for ILC detectors.

For the ILC study the following example point has been used:
\begin{eqnarray}
 m_{\neutralino{1}}       &=& 98.48 \GeV
  \label{eq:neutralinoinputmass}\\
 m_{\susy{\electron}_R}   &=& 280.72 \GeV
\end{eqnarray}
The mass of the neutralino has been selected as a worst case scenario, where $m_{\neutralino{1}}\simeq m_{\W/\Z}$. This is most
challenging, since in this case the LSP signal is expected to overlay significantly with SM background and the involved leptons from the
LSP decay to $\W$ become relatively soft. For higher LSP masses the study is almost SM background free. 
The production cross section for the polarisation $\mathcal{P}(\electron^+,\electron^-)=(-30\%,+80\%)$ and a center of mass energy of $\sqrt{s}=500\GeV$ amounts to $344\fb$, driven by the choice of $m_{\susy{\electron}_R}$. As can be seen from figure~\ref{fig:productioncrosssection}, the production
cross section is still $\sim 100\,$fb when $m_{\susy{\electron}_R}$ is twice as large.
The branching ratios for the decay modes that are relevant for measuring the neutrino atmospheric mixing angle (see 
eq.~\eqref{eq:theta23fromBR}) read\footnote{These branching ratios correspond to bRPV couplings $O_{\neutralino{1}Wl_{2,3}}$ in the order of $10^{-7}$.} $BR(\neutralino{1}\rightarrow\muon^{\pm}\W^{\mp})=0.43$ and $BR(\neutralino{1}\rightarrow\tauL^{\pm}\W^{\mp})=0.47$.
The remaining fraction comes mainly from on-shell $\Z$ decay modes $\neutralino{1}\rightarrow\neutrino_{i}\Z$, where the
 relative fraction of $\neutrino_{i}\Z$ vs.~$l_{i}\W$ decays
 depends only on the $\neutralino{1}$ mass. Decays to electrons have a negligibly small branching fraction due to the smallness of the reactor neutrino mixing angle $\theta_{13}$, since $\mathrm{BR}(\tilde{\chi}_1^0 \rightarrow W e)\propto\Lambda_e\propto\tan^2(\theta_{13})$~\cite{Porod:2000hv}. Three-body decays are negligible in our case. They  
 become sizable only for small $\susy{\tau}$-$\neutralino{1}$ mass differences, which would allow direct $\susy{\tau}$ production at
 the ILC and thus offer a wealth of additional observables which
 we don't discuss in this paper.
 
For an integrated luminosity of $100\ifb$ one expects $34400$ produced neutralino pairs and among them 
$6361$ events ending up in the $\muon\muon$-channel, $7599$ events in the  $\tauL\tauL$-channel, and $13904$ events in the mixed $\muon\tauL$-channel.
As the ILC is expected to deliver an integrated luminosity of $250\ifb/\mathrm{year}$ at $500\GeV$, this amount of data is going to be collected 
within approximately five months of operation at design luminosity.

\subsection{Data samples}
For the given example point a full detector simulation of the International Large Detector (ILD) based on the recently published detector descripton \cite{Behnke:2013lya} has been performed.
The ILD concept is one of two proposed detector concepts at the ILC.
Its design is optimised for Particle Flow reconstruction, which aims at reconstructing each individual particle with the most precise detector component.
To this end, properties of charged particles are only measured by the tracking system and the highly segmented calorimeters are only used for the measurement of neutral particles.
The main part of the proposed tracking system at the ILD is a time-projection-chamber, which is complemented by silicon strip and pixel detectors. This system is expected to obtain a tracking resolution of up to $\sigma_{1/p_t}=2\cdot10^{-5}\GeV^{-1}$ for high momenta.
Due to the more benign radiation environment at the ILC, the inner detectors can be built
with very little material, amounting to only $10\%$ of a radiation length in front of the electromagnetic calorimeter in the barrel region. The electromagnetic calorimeter is foreseen as a 30-layer silicon-tungsten sampling calorimeter with cell sizes of $5\times 5 \mm^2$. In testbeam operation of a prototype~\cite{bib:ecalperf}, an energy resolution of  $\Delta E/E = (16.6 \pm 0.1)\% / \sqrt{E} \oplus (1.1\pm 0.1)\%$ has been achieved.
The highly segmented hadronic calorimeter with cell sizes of $3\times3 \cm^2$ 
in connection with the Particle Flow Concept allows for a jet energy resolution of $\Delta E/E= (3-4)\%$  \cite{Thomson:2009rp}.

The full detector simulation has been performed for an integrated luminosity of $100\ifb$ and at a center of mass energy of $\sqrt{s}=500\GeV$.
For the SM background, samples produced for the benchmarking of the ILD detector for the Technical Design Report have been used~\cite{Behnke:2013lya}.

In the case of bRPV events the program {\tt Sarah}~\cite{Sarah} has been used to generate model files for the event generator.
As for the SM samples, {\tt Whizard}~\cite{Whizard} has been used as event generator of the hard process and {\tt Pythia}~\cite{Pythia} 
for fragmentation and hadronization. These events have been passed through \texttt{Mokka}~\cite{MoradeFreitas:2002kj}, the full 
{\tt Geant4}-based~\cite{Geant4} simulation of the ILD detector and finally reconstructed with {\tt MarlinReco}~\cite{Gaede:2006pj, bib:MarlinReco}.
For the event generation, realistic beam parameters have been taken into account,
in particular the ILC specific beam energy spectrum at $500\GeV$ \cite{Adolphsen:2013kya}.

With the instantaneous luminosity forseen at the ILC, on average $\langle N\rangle=1.2$ interactions of photons leading to the production of low $p_t$ hadrons are expected per bunch-crossing \cite{Behnke:2013lya}.
This takes into account contributions from real photons accompanying the electron beam due to bremsstrahlung and
synchrotron radiation as well as from virtual photons radiated off the primary beam electrons.
Therefore, each hard-interaction event (from SUSY or SM background) has been overlaid with a Poissonian distributed number of such $\gamma\gamma\rightarrow\mathrm{hadrons}$ events before the reconstruction step.

At the time of the MC production a too large number of overlay events per hard-interaction had been assumed ($\langle N\rangle=1.7$), which results in a conservative estimate of the $\gamma\gamma\rightarrow\mathrm{hadrons}$ background in this study. However, as it will be demonstrated in the next section, even this larger background could be removed very efficiently.

\subsection{Event selection}
The signal events have a rather clear signature: The produced LSPs decay into either a $\muon$ or $\tauL$ plus a $\W$ boson.
In the following we restrict ourselves to the hadronic $\W$ decay mode.
Thereby, the event is -- except for some missing energy from a potential $\tauL$ decay -- fully reconstructable with six visible objects
in the final state.

\subsubsection*{Event preparation and preselection}

Before the actual event selection is performed, the $\gamma\gamma\rightarrow\mathrm{hadrons}$ background is removed from the event
by the following procedure:
Since we expect to have six final state objects, an exclusive $k_T$ jet clustering algorithm ($R=1.3$) \cite{Cacciari:2011ma,hep-ph/0512210} which is forced to find six jets 
is applied to the reconstructed objects in the events.
This algorithm builds up six jets which are assigned to the hard interaction, 
as well as two very forward directed beam jets which are treated as beam background.
Removing those beam jets and using only the objects which end up clustered in the main jets for the further
analysis, recovers very well the bare event without background overlay. Fig.~\ref{fig:gagaremoval} shows the impact
of overlaid $\gamma \gamma$ events on the visible energy (left) and the ability to remove this background with the
described method (right).

\begin{figure*}
 \centering
 \includegraphics[width=0.495\textwidth]{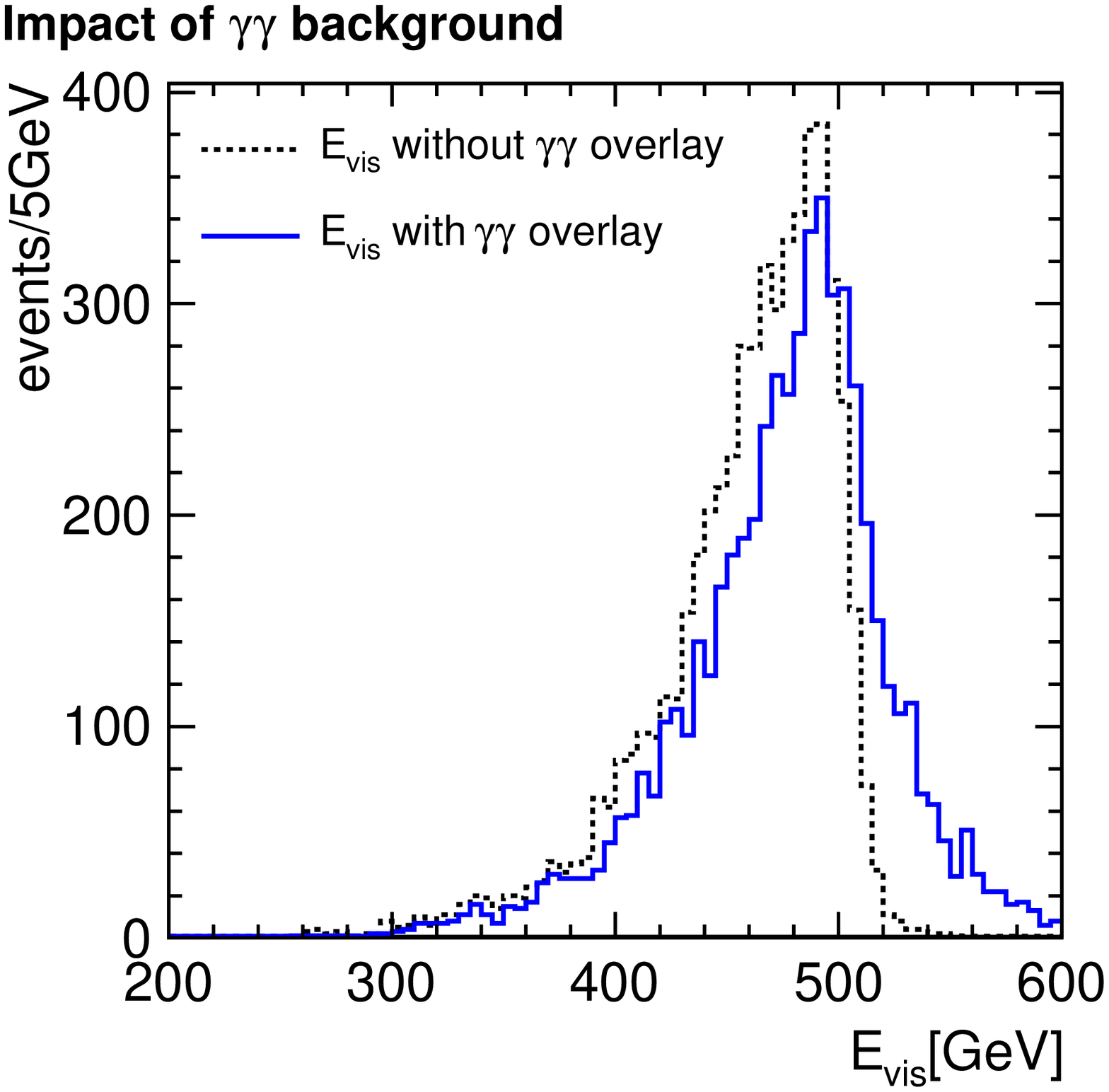}
 \includegraphics[width=0.495\textwidth]{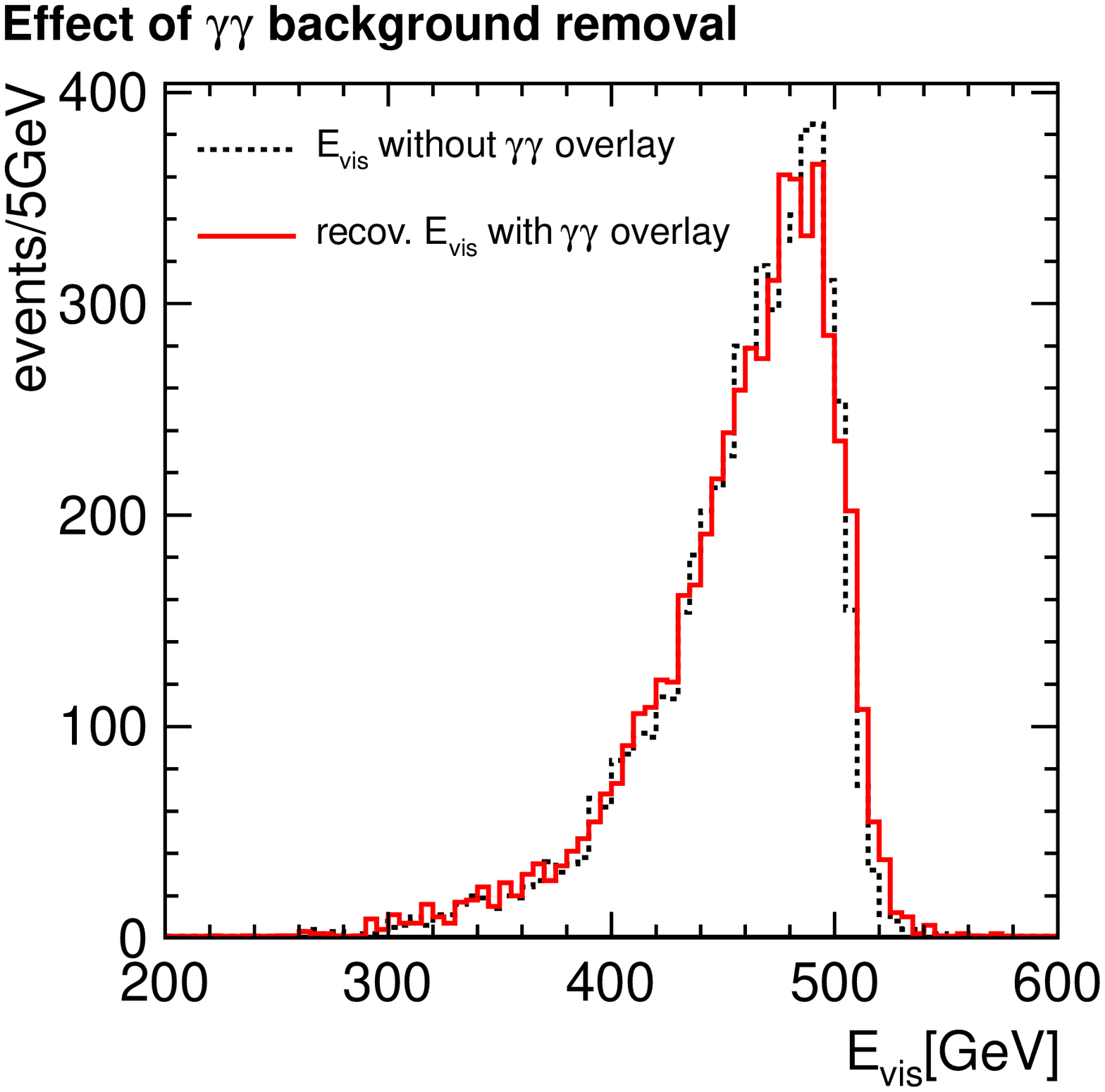}
 \caption{Left: Effect of $\gamma\gamma\rightarrow\mathrm{hadrons}$ background overlay on the visible energy in the bRPV SUSY sample.
          Right: Recovered visible energy in the event after $\gamma\gamma\rightarrow\mathrm{hadrons}$ background removal procedure described in the text.}
 \label{fig:gagaremoval}
\end{figure*}

Two hadronically decaying $\W$ bosons imply a relatively high particle multiplicity $N_{\mathrm{objects}}$ in the event.
Due to the fact that there is no major source for missing energy, the visible energy in the event is close to the center of mass energy of $500\GeV$.
So, the following preselection cuts have been used:
\begin{eqnarray}
 E_{\mathrm{visible}} \geq 350\GeV\\
 50 \geq N_{\mathrm{objects}} \geq 150,
\end{eqnarray}
This preselection on the one hand reduces Standard Model background and 
on the other hand cuts away some of the leptonic $\W$ contribution in the LSP decay, which is considered as background in this analysis. $95\%$ of the signal events pass this preselection.

In order to be able to measure the ratio of different branching ratios of the LSP decay (see eq.~\eqref{eq:theta23fromBR})
we have to define a selection to distinguish between the different event classes:

\subsubsection*{$\mathbf{\muon\muon}$ class}
Here we look for an event with at least two reconstructed muons. The muon identification is provided by the Pandora Particle Flow Algorithms \cite{Thomson:2009rp}, seeded by a minimum-ionizing signature in the calorimeters and the instrumented flux return yoke.
The two most energetic muons are removed from the event.
  No further lepton isolation criteria is required. The remaining objects of the event are clustered by the Durham jets clustering algorithm
  into four jets. All pair-permutations of the four jets are used to find the best $\W$ candidates, where the objects with the smallest
  \begin{eqnarray}
   \chi^2_{\W_i}=\left(\frac{m_{\mathrm{reco,}i}-m_{\W}}{\sigma_{\mathrm{res}}}\right)^2
  \end{eqnarray}
  are used for the proceeding analysis.
  Herein, $\sigma_{\mathrm{res}}$ is an estimated resolution factor of $5\GeV$.
  For all combinations of $\W$ candidates and muons the invariant mass is determined and the reconstructed objets from the pair with smallest
  \begin{eqnarray}
  \chi^2_{\mathrm{eqm}}=\left(\frac{m_{\mathrm{reco,}1}-m_{\mathrm{reco,}2}}{\sigma_{\mathrm{res}}}\right)^2
  \end{eqnarray}
  are considered as LSP candidates.
  If the event fulfills the condition 
  \begin{eqnarray}
   \chi^2_{\W_1}<2 \quad \mathrm{and} \quad \chi^2_{\W_2}<2 \quad \mathrm{and} \quad \chi^2_{\mathrm{eqm}}<2,
   \label{eq:eventclasscondition}
  \end{eqnarray}
  the event is counted as $\muon\muon$ event, else it is tested against the $\muon\tauL$ class.

\subsubsection*{$\mathbf{\muon\tauL}$ class}
 In this class at least one reconstructed muon in the event is required. The most energetic muon is removed 
  and the rest of the event is forced into five jets by an inclusive Durham jet clustering algorithm.
  Since the muon requirement is relaxed, this class is more prone to background. Therefore we test in addition
  whether the  chosen jet configuration describes the event well. We employ a cut on the Durham jet algorithm parameters $y_{i-1,i}$ and $y_{i,i+1}$ \cite{Cacciari:2011ma}.
  $y_{i-1,i}$ is a measure of the distance in energy-momentum space between the two of the $i$ jets which would be merged  if a $i-1$ configuration
  was required. $y_{i,i+1}$ gives the analoguos distance measure for the two jets which have been merged in the last step when clustering from the $i+1$
  to the final $i$ configuration.
  For events well fitting to the 5-jet configuration, $y_{4,5}-y_{5,6}$ gets maximal, so we use the following cut:
  \begin{eqnarray}
  y_{4,5}-y_{5,6}>5\cdot10^{-5}   \label{eq:jetyparametercut}.
  \end{eqnarray}
  The jet with the smallest number of constituents is considered as $\tau$ candidate if $N_{\mathrm{const.}}<10$.
  Additionally it is required that the jet does not contain a muon.
  The further approach to find $\W$ candidates and finally LSP candidates is identical to the $\muon\muon$ class.
  If condition \eqref{eq:eventclasscondition} is satisfied this event is counted as $\muon\tau$ event, otherwise the $\tauL\tauL$ class is tested.

\subsubsection*{$\mathbf{\tauL\tauL}$ class}
 The event is forced into six jets by an inclusive Durham jet clustering algorithm. In analogy to the $\muon\tauL$ class,
  it is required that $y_{5,6}-y_{6,7}>5\cdot10^{-5}$.
  The two jets with the smallest numbers of constituents are considered as $\tau$ candidates if each fulfills
  $N_{\mathrm{const.}}<10$ and the muon-veto.
  The further approach to find $\W$ candidates and finally LSP candidates is again identical to the $\muon\muon$ and $\muon\tauL$ class.
  If condition \eqref{eq:eventclasscondition} is satisfied this event is counted as $\tau\tau$ event. Otherwise, the event is rejected.

\begin{figure*}
 \centering
 \includegraphics[width=0.32\textwidth]{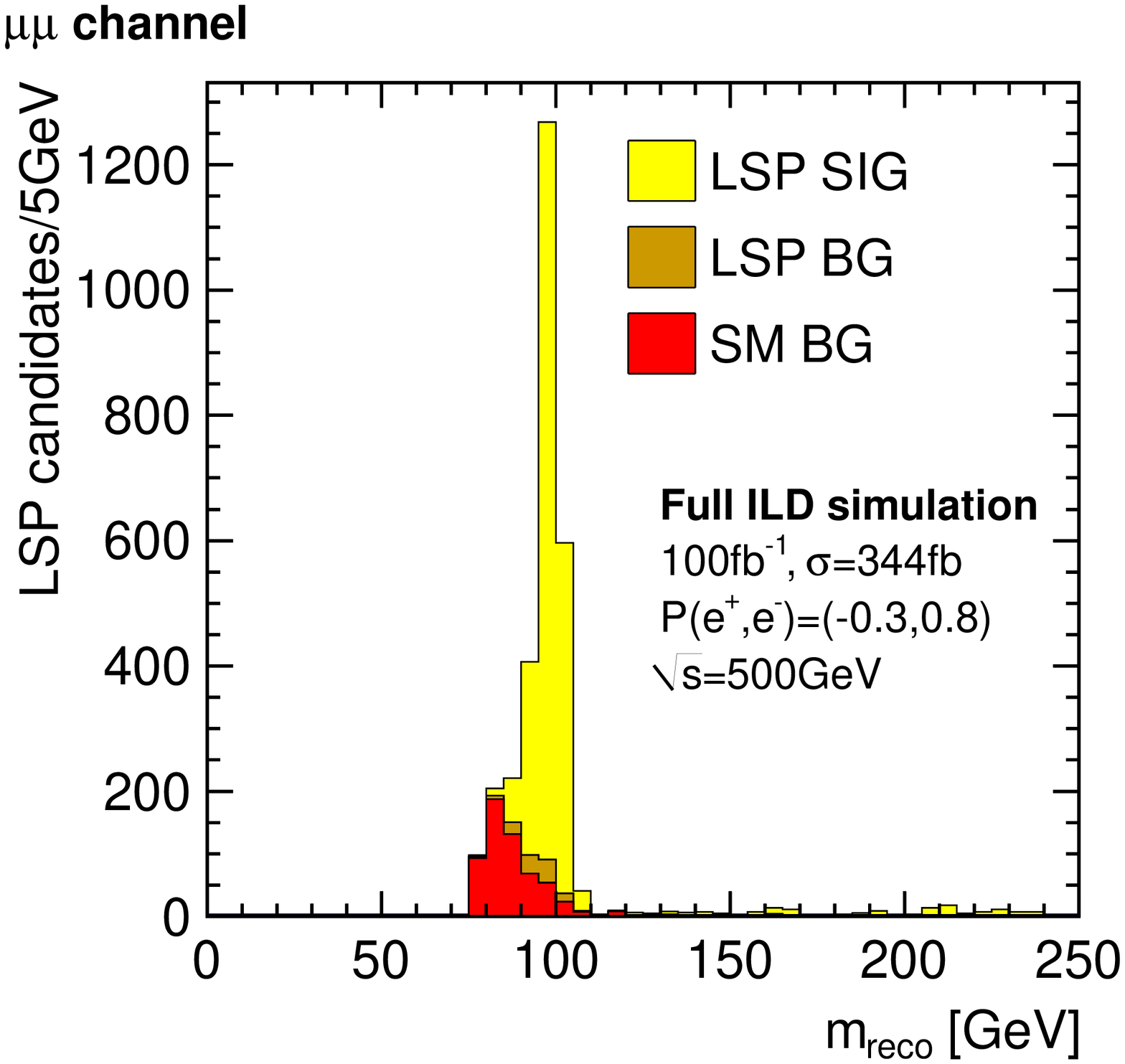}
 \includegraphics[width=0.32\textwidth]{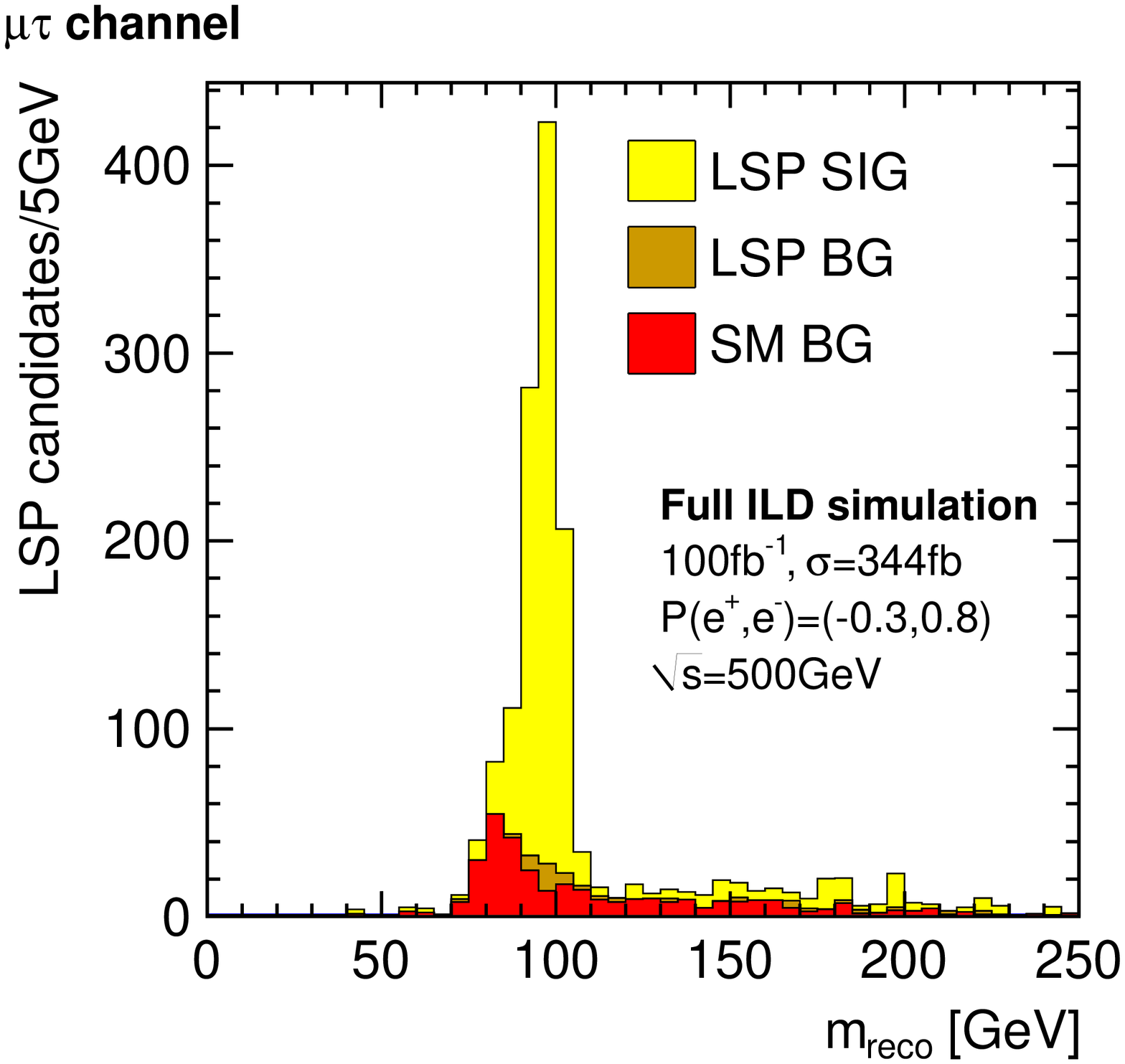}
 \includegraphics[width=0.32\textwidth]{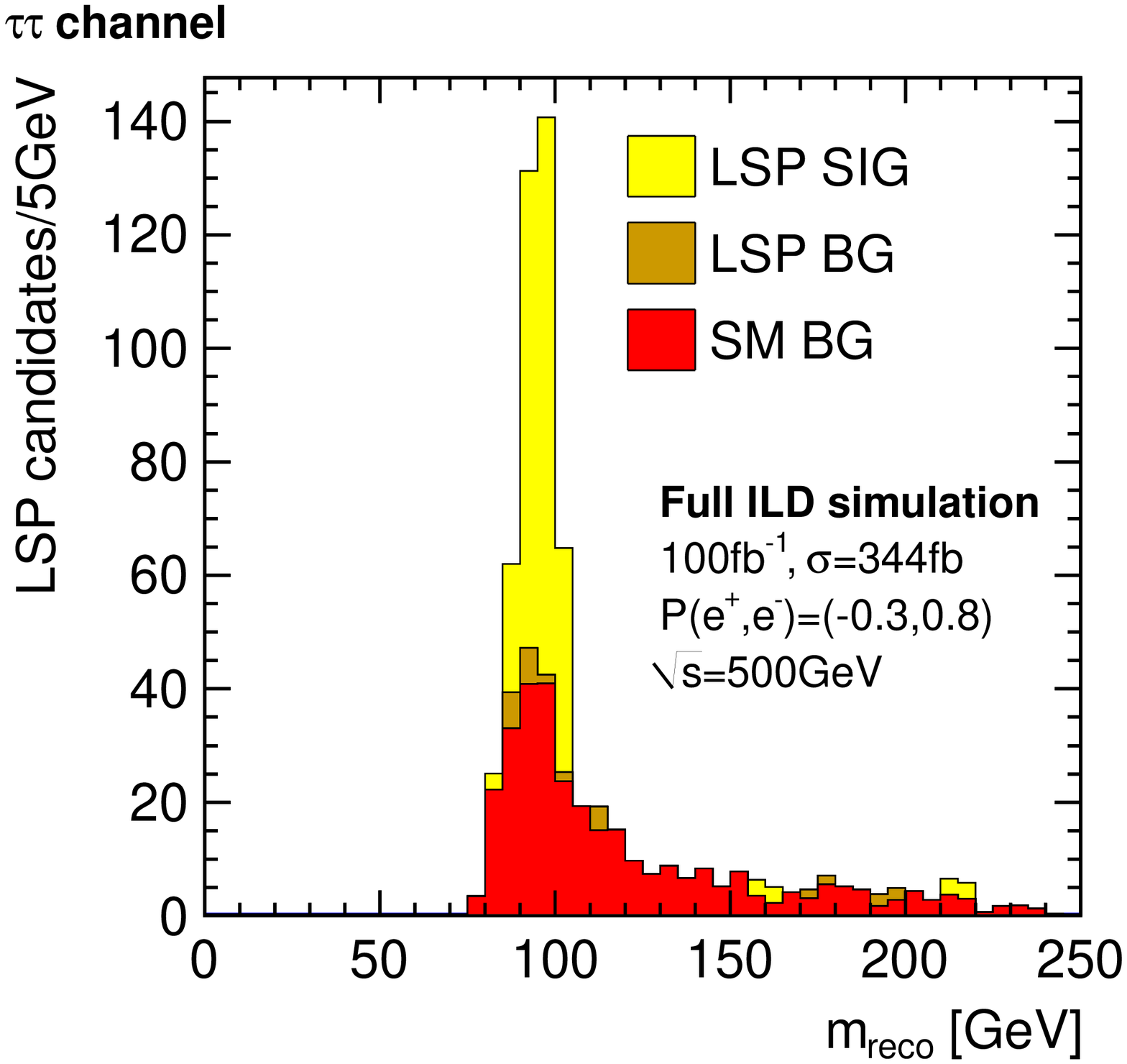}
 \caption{Reconstruction mass of the LSP for the different event classes. Yellow indicates the signal events, orange depicts background originating from non-signal LSP decays and
  red shows the remaining Standard Model background. For the $\mu\mu$ and $\mu\tau$ channel the SM background peaks at $M_W/M_Z$ and the signal at $m_{\neutralino{1}}$.
  In the $\tau\tau$ channel the purity is lowest.}
 \label{fig:massreconstruction}
\end{figure*}

After this selection fig.~\ref{fig:massreconstruction} shows the reconstructed neutralino mass in each of the event classes for $100\ifb$ of fully simulated events. 
The remaining background is dominated by SM $W$ pair production, whereas the
contribution from other LSP decays is very small.

\subsection{Systematic uncertainties} \label{subsec:systematicuncertainties}
\subsubsection*{Mass reconstruction}
For the reconstruction of the LSP mass the two most important sources of uncertainties arise from the reconstruction of the $\muon$ momentum as well as on the reconstruction of the momentum and energy of the $\W$ candidates. In our scenario, the muons originating from an LSP decay carry a momentum of up to $80\GeV$, while the jets originating from the $\W$ bosons from the signal decay have an energy of up to $200\GeV$.

The muon momentum scale can be calibrated using $Z$ boson decays. The unpolarised cross section of the process $\electron^+\electron^-\rightarrow Z \rightarrow\muon^+\muon^-$ at the ILC500 amounts to $2.5\pb$. For an assumed integrated luminosity of $\intL=500\ifb$ this yields in total $N_{\muon\muon}=12.5\cdot10^5$ muon pairs. Assuming that ultimatively the precision of the scale calibration is limited by the available statistics
for the calibration process, we estimate that a precision of $1/\sqrt{N_{\mu\mu}}=0.09\%$ can be reached.

The jet energy calibration of hadronically decaying $\W$ bosons can be derived 
analoguously from hadronic $\Z$ decays. Since the cross section for $\electron^+\electron^-\rightarrow Z \rightarrow \quark\bar{\quark}$ is ten times larger compared to the muonic decay channel, which means that the from the point of view of available control sample statistics, the jet energy scale uncertainty 
could reach $0.03\%$, assuming that the calibration is sufficiently stable over time. The resulting impact on the neutralino mass determination is $11\MeV$. Thus it could
be a factor four larger before it becomes comparable to the contribution from the momentum scale. Alternatively, exploiting kinematic fits~\cite{Beckmann:2010ib} in conjunction with the well-known beam energy, at the ILC forseen to be controlled to  $10^{-4}$~\cite{Behnke:2013lya}, could significantly reduce the dependence
of the reconstructed neutralino mass on the jet energy scale, leading to similar final
precision estimates.

Systematic errors on luminosity, beam energy and beam polarisation do not enter into the mass measurement, which depends solely on the reconstructed detector signals. The selection efficiency does not show any dependency on the reconstructed LSP mass within the available Monte-Carlo statistics. Thus we conclude that any potential bias due to the selection efficiency plays an insignificant role in the mass measurement.

For LSP masses below $\sim105\GeV$, the SM background has a steeply falling invariant mass
distribution. Therefore any uncertainty related to the modelling of this slope, e.g. assumptions on hadronisation, colour-reconnection etc., could enter into the LSP mass determination. However the ILC itself will offer numerous opportunities for SM precision measurements beyond today's knowledge. In particular the cross section for the dominating background process, $W$ pair production, is in the several pb range, thus providing ample
possibilities to tune the modelling of this process. We thus estimate that the residual 
effect on the LSP mass determination via subtraction of the SM background is not larger
than $15\MeV$.

\begin{table}
\centering
\begin{tabular}{|c|c|c|c|c|}
\hline
 source                    & calibration process           & cross section  & $1/\sqrt{N}$@$500\ifb$ & effect on $m_{\neutralino{1}}$ \\ \hline
 $|\vec{p}_{\muon}|$ scale & $\Z\rightarrow \muon^+\muon^-$    & $2.5\pb$       & $0.09\%$       & $46\MeV$ \\
 $E_{\mathrm{jet}}$ scale  & $\Z\rightarrow \quark\bar{\quark}$  & $25\pb$        & $0.03\%$       & $11\MeV$ \\
 background modelling & $\W\W \rightarrow$ hadrons & $7\pb$ & $0.05\%$ & $15\MeV$ \\
 \hline
 total & & & & $\sim50\MeV$ \\
\hline
\end{tabular}
\caption{Main sources for systematic uncertainties on the LSP mass reconstruction. The largest contribution in the current estimate stems from the muon momentum scale calibration.}
\label{tab:systuncertmass} 
\end{table}

Table \ref{tab:systuncertmass} summarises the main sources for the systematic uncertainty on the LSP mass reconstruction and its propagation to the LSP mass reconstruction.
 The total systematic error is estimated to be about $50\MeV$.

\subsubsection*{Measurement of ratio of branching ratios}
The measurement of the ratio of branching ratios is a measurement of the ratio of the number of events reconstructed in the different event classes.
For this reason, all systematic uncertainties which factorise with the number of events cancel.
The same is true for reconstruction effects which affect all event classes simultanousely, like systematic uncertainties on jet energy scales, for instance.

The main source of systematic error is expected to arise from the determination of the selection efficiencies and purities of the different event classes from Monte-Carlo.
Therefore, validation of the Monte Carlo simulation with data is very important.

The process $\electron^+\electron^-\rightarrow \Z\Z \rightarrow\lepton^+\lepton^- \quark \bar{\quark}$ with $l=\{\muon,\tauL\}$ offers the possibility to study the Monte-Carlo description of $\tau$ + jets and $\mu$ + jets events under comparable experimental conditions as for the signal decay.
The unpolarised cross section per process at a center of mass energy of $500\GeV$ is $225\fb$. Thus, for $\intL = 500\ifb$ the expected precision on Monte-Carlo and data comparison amounts to $0.3\%$.
As a second process, $\electron^+\electron^-\rightarrow \W^+\W^- \rightarrow\lepton \neutrino_{\lepton} \quark \bar{\quark}$ with $l=\{\muon,\tauL\}$ can also be studied in order to validate Monte-Carlo. This process has a significantly larger cross section of $2.5\pb$ per process, which leads in the end to a statistical uncertainty of $0.09\%$ for $\intL = 500\ifb$.

As for the mass measurement, we allow some modelling uncertainty for the main SM background, $\W\W\rightarrow$ hadrons, of $0.05\%$.

Taking these considerations into account, a conservative estimate on the systematic uncertainties on Monte-Carlo is $\mathcal{O}(0.5\%)$ (c.f. table \ref{tab:systuncerteffi}).
\begin{table}
\centering
\begin{tabular}{|c|c|c|c|}
\hline
            & control process                                               & cross section  & $1/\sqrt{N}$@$500\ifb$\\ \hline
  \multirow{2}{*}{$\muon$ / $\tau$ ID with  $\Z/\W \rightarrow$ jets }    & $\Z\Z\rightarrow \lepton\lepton \quark\bar{\quark}$  & $225\fb$       & $0.3\%$\\
                & $\W\W\rightarrow \lepton \neutrino_{\lepton} \quark \bar{\quark}$ & $2.5\pb$       & $0.09\%$\\
 SM background modelling & $\W\W \rightarrow$ hadrons & $7\pb$ & $0.05\%$ \\
 \hline
 total & & &  $\sim 0.5\%$ \\
\hline
\end{tabular}
\caption{Possible control samples to verify efficiencies and purities obtained from Monte-Carlo simulation; $\lepton$ denotes either a $\muon$ or a $\tau$. The largest contribution originates from the limited statistics of the SM control sample for verifying
the migrations between the $\muon\muon$, $\muon\tau$ and $\tau\tau$ classes.}
\label{tab:systuncerteffi} 
\end{table}

%
\section{Results} \label{sec:results}
\subsection{Mass measurement and resolution}
The $\muon\muon$ channel shows a clear signal peak, which can be used for measuring the LSP mass accurately. Figure~\ref{fig:precisionreco} shows the reconstructed LSP mass spectrum for an integrated luminosity of $\intL=100\ifb$ after subtracting the SM background. Since the natural width of the LSP is negligibly small ($\Gamma_{\neutralino{1}}=\mathcal{O}(10^{-14}\GeV$)), the width of the distribution is dominanted by the detector resolution. From a gaussian fit, the LSP mass can be 
determined to
\begin{align}
m^{\mathrm{fit}}_{\neutralino{1}}&=(98.401\pm0.092\text{(stat.)})\GeV .
\end{align}
This value is within the error in very good agreement with the input mass of the example point of $98.48\GeV$ (c.f. eq.~\eqref{eq:neutralinoinputmass}). The obtained width of the gaussian is about $\sigma^{\mathrm{fit}}_{\neutralino{1}} = 3\GeV$, which is in very good agreement with the ILD design goals~\cite{ILDLoI}.

Scaling the statistical uncertainty to an integrated luminosity of $\intL=500\ifb$ and combining the systematic uncertainties as discussed in section \ref{subsec:systematicuncertainties}, the total uncertainty on the LSP mass measurement becomes
\begin{align}
\delta m^{\mathrm{fit}}_{\neutralino{1}}&=(40\text{(stat.)} \oplus 50\text{(syst.)})\MeV.
\end{align}

\begin{figure}
 \centering
 \includegraphics[width=0.495\textwidth]{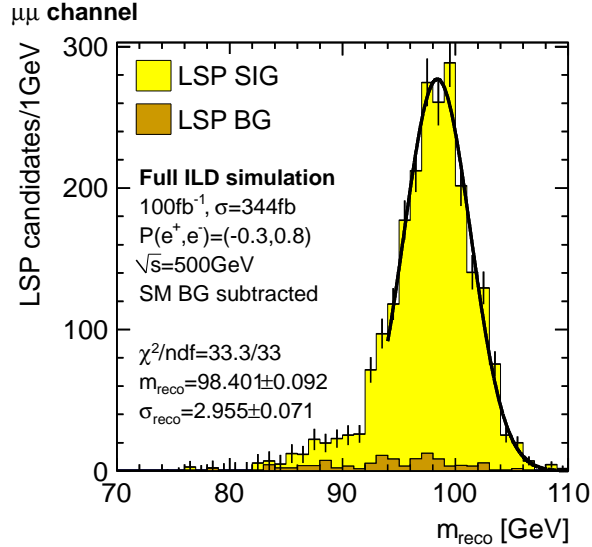}
 \caption{Mass reconstruction of the LSP in the $\muon\muon$ channel. The Standard Model background is subtracted, but fluctuations are taken into account in the bin errors. The distribution is slightly washed out towards lower masses, which originates from misreconstructions. For this reason, this part of the distribution has not been taken into account for the fit.}
 \label{fig:precisionreco}
\end{figure}%

\subsection{Signal significance}
The precision measurement of the LSP mass in the $\muon\muon$ channel can be used to define a signal region
$m^{\mathrm{fit}}_{\neutralino{1}}\pm3\sigma^{\mathrm{fit}}_{\neutralino{1}}$. This further reduces the background fraction in the selected event classes.
The decomposition of number of measured events in the different event classes $N^{\mathrm{reco}}$ into the number of events
in the different truth classes $N^{\mathrm{true}}$ is shown in the the following matrix $\mathbf{N}$:
\begin{eqnarray}
\label{eq:eventmatrix}
\mathbf{N}=\bordermatrix{
 & N^{\mathrm{true}}_{\muon\muon} & N^{\mathrm{true}}_{\muon\tauL} & N^{\mathrm{true}}_{\tauL\tauL} & N^{\mathrm{true}}_{\mathrm{LSP BG}} & N^{\mathrm{true}}_{\mathrm{SM BG}} \cr
N^{\mathrm{sel}}_{\muon\muon} & 858 & 173 &  11 & 40 & 69 \cr
N^{\mathrm{sel}}_{\muon\tauL} & 16  & 410 &  45 & 17 & 67 \cr
N^{\mathrm{sel}}_{\tauL\tauL} & 0   &   2 & 107 &  4 & 60 \cr
}
\end{eqnarray}
Thereby, $N^{\mathrm{true}}_{\mathrm{LSP BG}}$ counts events in which at least one LSP decays differently than the targeted two-body
decay $\neutralino{1}\rightarrow\W\lepton$.
As expected, the $\tauL\tauL$ channel is by far the one with the lowest purity.

The signal over Standard Model background ratio for the different classes can be derived.
Herein, every selected event originating from an LSP decay is counted as signal.
\begin{center}
\begin{tabular}{lcc}
                   &  $S/\sqrt{B}$ \\
$\muon\muon$ class &  130 \\
$\muon\tauL$ class &  60 \\
$\tauL\tauL$ class &  15 
\end{tabular}
\end{center}

The described event selection is still significant on a $5\sigma$ level for very large selectron masses well above $1.5\TeV$ and a large range of $m_{\neutralino{1}}$, as depicted in fig.~\ref{fig:discoverypot}. Positron beam polarisation further enhances the production cross section and, thus, increases the sensitivity of the analysis to selectron
masses of almost $2\TeV$ for $P(e^+)=-60\%$.

\begin{figure}
 \centering
 \includegraphics[width=0.495\textwidth]{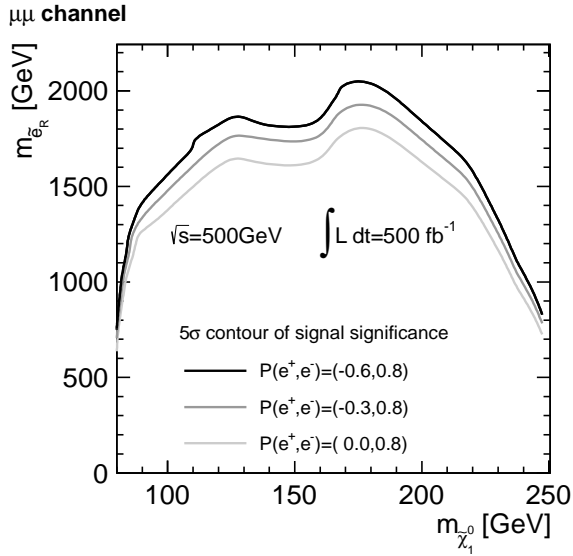}
 \caption{5$\sigma$ contour of signal significance in the $\mu\mu$ channel for different beam polarisations and an integrated luminosity of $\intL = 500\ifb$ based on a log likelihood ratio. The di-muon selection efficiency obtained from full simulation for $m_{\neutralino{1}}=98.48\GeV$ has been assumed for all values of the neutralino mass. For larger $m_{\neutralino{1}}$, this is a conservative estimate. Higher positron beam polarisation helps to further increase the significance in this model.}
 \label{fig:discoverypot}
\end{figure}%

It has already been pointed out that this studied parameter point is a worst case scenario with respect to the neutralino mass.
It is clearly visible in fig.~\ref{fig:massreconstruction} that for another parameter point with a higher LSP mass the signal peak would shift into an
almost completely background-free region.
Though, the LSPs in this model become rather long-lived at lower masses (compare fig.~\ref{fig:xseclength}), where the Standard Model background is large.
Adding this information to the analysis and requiring from the reconstructed objects not to point to
the primary vertex, would reduce the Standard Model background drastically.
However, this would on the other hand introduce a strong model dependency to the analysis. Therefore the exploitation of the lifetime information is left as a future option for improvements.

Likewise, the requirement for same sign leptons in the event classes is an option for further improving the analysis.
This restriction would reduce the number of signal events by a factor of $2$, but could heavily suppress the remaining SM background.

\subsection{Branching ratio measurement}
\label{sec:brmeasurement}
We assume in the following that the average number of Standard Model background events can be estimated from Monte Carlo with a precision of $0.05\%$ (cf. section \ref{subsec:systematicuncertainties}).
The LSP non-signal background consists mainly of events in which one of the two LSPs decayed non-signal like into $\Z\neutrino$.
As soon as the LSP mass is known, the relative fraction $\BR(\neutralino{1}\rightarrow\Z\neutrino)/\BR(\neutralino{1}\rightarrow\W\lepton)$ is determined and, thus, also the number of LSP background events can be predicted.
Under this assumption we can subtract the backgrounds and build a $3\times3$ efficiency matrix $\mathbf{E}$, which is defined like
\begin{eqnarray}
(\mathbf{E})_{ij}=\frac{(\mathbf{N})_{ij}}{N^{\mathrm{true}}_j}=\left(
\begin{array}{ccc}
 0.2981     & 0.0277   & 0.0032 \\
 0.0056     & 0.0658   & 0.0129 \\
 0.0000     & 0.0004   & 0.0306  
\end{array}
\right)_{ij},
\end{eqnarray}
where $i,j={\mu\mu,\mu\tau,\tau\tau}$.
The error on the entries of the efficiency matrix is dominated by the assumed systematic uncertainty of $0.5\%$ on the Monte-Carlo prediction on the migrations between the signal classes (cf. section \ref{subsec:systematicuncertainties}).

The vector of selected events becomes
\begin{eqnarray}
\vec{N}_{\mathrm{sig}}^{\mathrm{sel}}&=
 \left(
\begin{array}{c}
N^{\mathrm{sel}}_{\muon\muon}-\langle N^{\mathrm{MC}}_{BG,\muon\muon}\rangle\\
N^{\mathrm{sel}}_{\muon\tauL}-\langle N^{\mathrm{MC}}_{BG,\muon\tauL}\rangle\\
N^{\mathrm{sel}}_{\tauL\tauL}-\langle N^{\mathrm{MC}}_{BG,\tauL\tauL}\rangle
\end{array}
\right)\\
\Delta\vec{N}_{\mathrm{sig}}^{\mathrm{sel}}&=
\left(
\begin{array}{c}
\sqrt{N^{\mathrm{sel}}_{\muon\muon}+\delta\langle N^{\mathrm{MC}}_{BG,\muon\muon}\rangle ^2}\\
\sqrt{N^{\mathrm{sel}}_{\muon\tauL}+\delta\langle N^{\mathrm{MC}}_{BG,\muon\tauL}\rangle ^2}\\
\sqrt{N^{\mathrm{sel}}_{\tauL\tauL}+\delta\langle N^{\mathrm{MC}}_{BG,\tauL\tauL}\rangle ^2}
\end{array}
\right).
\end{eqnarray}
Herein, $\delta\langle N^{\mathrm{MC}}_{BG,\lepton\lepton}\rangle$ with $\lepton\lepton=\{\muon\muon, \muon\tauL, \tauL\tauL\}$ are the systematic errors on the SM background estimation in the event classes.
They are negligible compared to the statistical fluctuations of the reconstructed events per event class.

The efficiency matrix can then be inverted and used to unfold the different event classes obtaining the number of reconstructed events in the event classes
\begin{eqnarray}
\vec{N}^{\mathrm{reco}}= \mathbf{E}^{-1} \vec{N}^{\mathrm{sel}}_{\mathrm{sig}}.
\end{eqnarray}

The ratio of the two branching rations can be extracted in different ways
\begin{eqnarray}\label{eq:relations}
 \frac{\BR(\neutralino{1}\rightarrow\W\mu)}{\BR(\neutralino{1}\rightarrow\W\tau)} &= \frac{2N^{\mathrm{reco}}_{\muon\muon}}{N^{\mathrm{reco}}_{\muon\tauL}}
 = \frac{N^{\mathrm{reco}}_{\muon\tauL}}{2N^{\mathrm{reco}}_{\tauL\tauL}}
 = \sqrt{\frac{N^{\mathrm{reco}}_{\muon\muon}}{N^{\mathrm{reco}}_{\tauL\tauL}}},
\end{eqnarray}
since for the expected number of events the following relations hold:
\begin{eqnarray}
 N^{\mathrm{reco}}_{\muon\muon}&=&N_{\neutralino{1}\neutralino{1}}\cdot\BR^2(\neutralino{1}\rightarrow\W\muon)\\
 N^{\mathrm{reco}}_{\muon\tauL}&=&N_{\neutralino{1}\neutralino{1}}\cdot2\cdot\BR(\neutralino{1}\rightarrow\W\muon)\cdot\BR(\neutralino{1}\rightarrow\W\tauL)\\
 N^{\mathrm{reco}}_{\tauL\tauL}&=&N_{\neutralino{1}\neutralino{1}}\cdot\BR^2(\neutralino{1}\rightarrow\W\tauL),
\end{eqnarray}
where $N_{\neutralino{1}\neutralino{1}}=\sigma(\electron^{+}\electron^{-}\rightarrow\neutralino{1}\neutralino{1})\cdot\intL$ is the number of produced LSP pairs.
Because of the low selection purity in the $\tauL\tauL$ channel, we have chosen the
relation involving only the $\muon\muon$ and $\muon\tauL$ channel for the further analysis.
Using all three relations in eq. \eqref{eq:relations} as input for a constrained fit can improve the precision by a factor of $2$, but this is not persued in the following.

In order to estimate the uncertainty on the resulting ratio of event numbers, an error propagation has been 
performed.
The uncertainty is depicted in fig.~\ref{fig:brprecision} in dependence of the integrated luminosity at the ILC500.
For the studied parameter point the achievable precision for $\intL=100\ifb$ is about $9\%$. This scales down to 
roughly $4\%$ for $500\ifb$, which is the desired integrated luminosity at ILC500 in a first stage.
The uncertainty contains a systematic error of $0.85\%$ arising from the propagation of the assumed systematic uncertainties on the efficiencies~\ref{tab:systuncerteffi} through the unfolding procedure.

\begin{figure}
 \centering
 \includegraphics[width=0.475\textwidth]{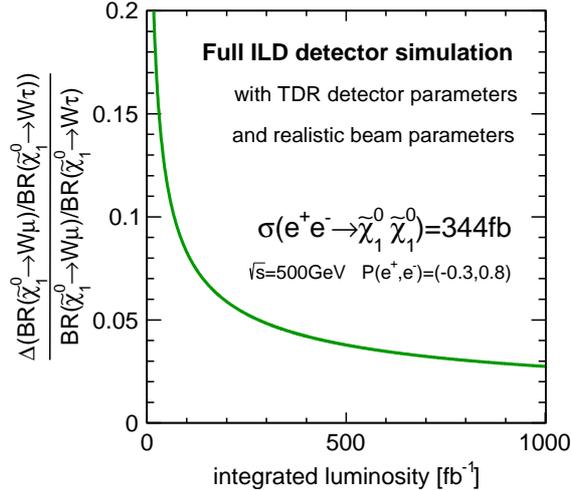}
 \caption{Achievable measurement precision of the ratio of the muonic and tauonic two-body decay modes of the LSP at the ILC.}
 \label{fig:brprecision}
\end{figure}

\subsection{Neutrino interpretation}
\label{sec:neutrinointerpretation}

The measured ratio of branching ratios can now be translated into the atmospheric mixing angle following eq.~\eqref{eq:theta23fromBR}.
As mentioned earlier, the given relation is only valid on tree-level and there are additional parametric uncertainties coming
from residual SUSY parameter dependencies \cite{Porod:2000hv}.
For this reason we define two scenarios: In the first scenario we assume that the LSP is the only accessible SUSY particle at ILC500. All other supersymmetric particles are randomly chosen to be heavier than $300\GeV$. In a random parameter scan consisting of $~6000$ scan points \cite{Porod:2013} we find that for $95\%$ of all found viable SUSY parameter points the deviation of the correlation between atmospheric mixing angle and the ratio of the branching ratios is below $17\%$. Assuming that the remaining three electroweakinos are measureable at ILC1000, which is an optional upgrade of the ILC to a center of mass energy of $1\TeV$, the correlation uncertainty reduces to $7\%$.

\begin{figure}
 \centering
 \includegraphics[width=0.4298\textwidth]{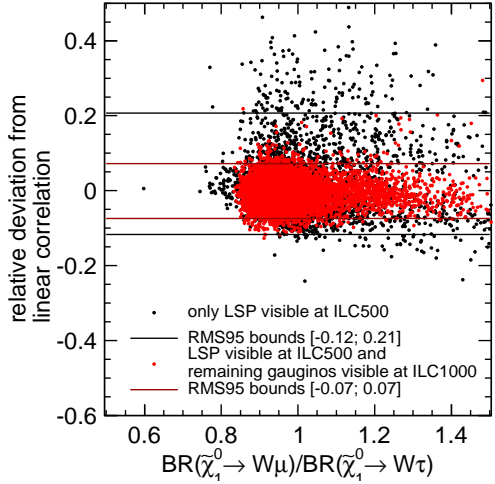}
 \caption{Random scan of the SUSY parameter space \cite{Porod:2013} in order to estimate the parametric uncertainty on the correlation between the atmospheric mixing angle and the ratio of branching ratios $\BR(\neutralino{1}\rightarrow\W\mu)/\BR(\neutralino{1}\rightarrow\W\tau)$.}
 \label{fig:parametricuncert}
\end{figure}

The derived precision of the measurement of the atmospheric neutrino mixing angle for different assumed parametric uncertainties
is shown in fig.~\ref{fig:brfinalprecision} (left).
One can now compare this precision with the uncertainty of current neutrino experiments \cite{Fogli:2012ua}, which is done in fig.~
\ref{fig:brfinalprecision} (right). The middle red line indicates the best fit value of the atmospheric neutrino mixing angle and
the upper and lower dashed red lines indicate the $1\sigma$ uncertainty. 
An agreement between the collider and neutrino experiment data would clearly
establish bRPV as origin of neutrino masses. Improvements from future neutrino 
experiments or a reduction of the parametric uncertainty by observation of additional SUSY particles at the ILC or the LHC would strengthen this conclusion even further.

\begin{figure*}
 \centering
 \includegraphics[width=0.475\textwidth]{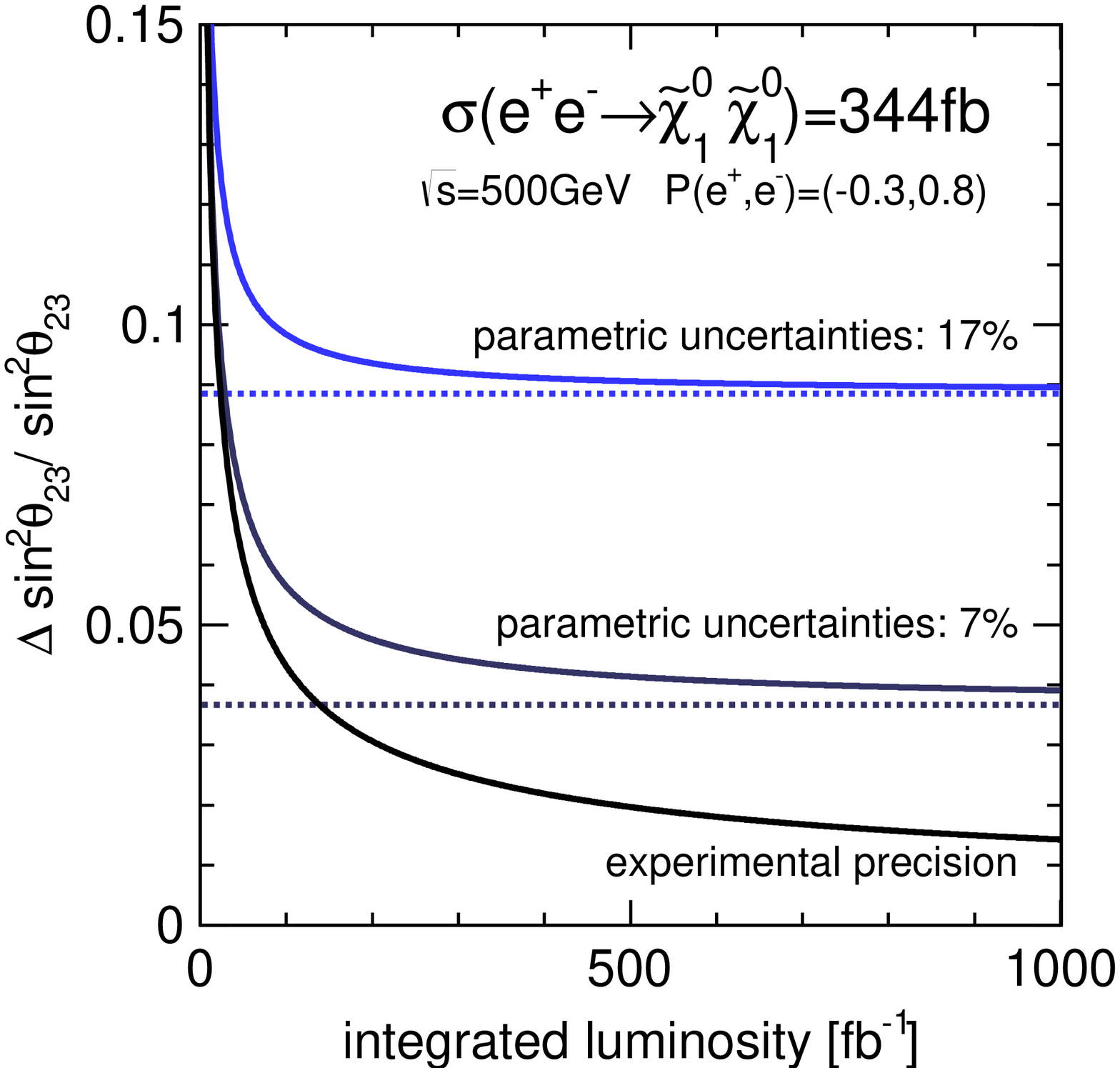}
 \includegraphics[width=0.475\textwidth]{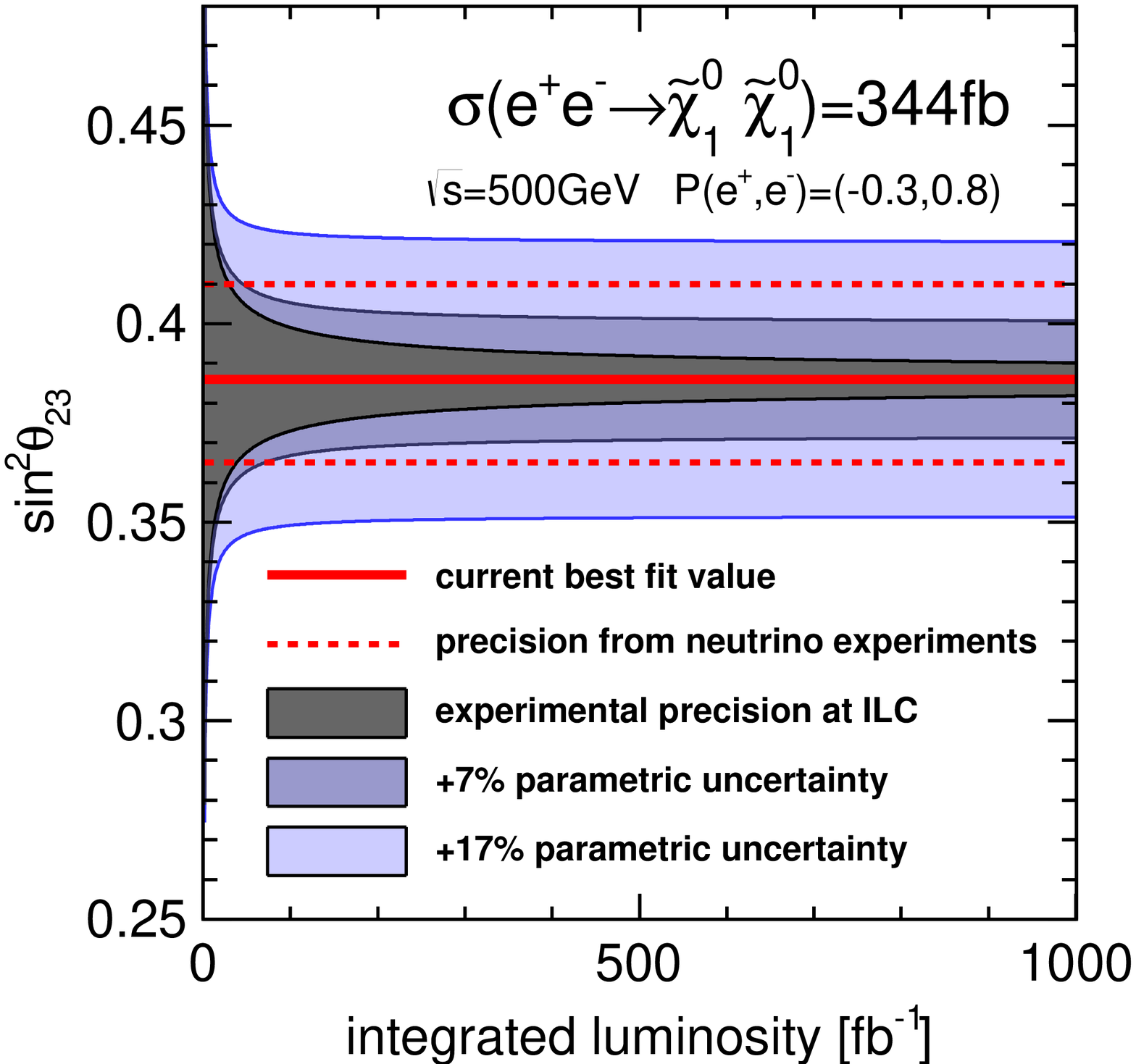}
 \caption{Precision of the measurement of the atmospheric mixing angle at the ILC. Left: Relative uncertainty assuming different 
 parametric uncertainties on the relation between ratio of branching rations and atmospheric neutrino mixing angle.
 Right: Comparison between achievable precision at the ILC and the precision at current neutrino experiments assuming present best fit value \cite{Fogli:2012ua} as central value.}
 \label{fig:brfinalprecision}
\end{figure*}

%
\section{Conclusions} \label{sec:conclusions}
We have presented a full ILD detector simulation of a bRPV SUSY model, which is an attractive
possibility to explain neutrino mass generation and mixing.
A highly detailed ILD model as well as realistic ILC beam parameters have been taken into account for the simulation.
As studied parameter point a worst case scenario has been used, where $m_{\neutralino{1}}\simeq m_{\W/\Z}$ and, thus, the signal
significantly overlaps with SM background.

We have developed a model-independent selection strategy to disentangle the different event classes involving the 
two decay modes of the LSP $\neutralino{1}\rightarrow\muon^{\pm}\W^{\mp}$ and $\neutralino{1}\rightarrow\tauL^{\pm}\W^{\mp}$.
It has been demonstrated that in the $\muon\muon$ event class a very accurate mass measurement
with an uncertainty of $\delta m^{\mathrm{fit}}_{\neutralino{1}}=(40\text{(stat.)} \oplus 50\text{(syst.)})\MeV$ is possible for an integrated luminosity of $500\ifb$.
With the described selection, a  signal to background ratio of $130$ in the $\muon\muon$ event class and $60$ in
the $\muon\tauL$ event class has been achieved.
Even for very large selectron masses of up to $1.5\TeV$ a $5\sigma$ discovery is possible for a large range of $m_{\neutralino{1}}$.

The $\muon\muon$ and $\muon\tauL$ event class have been used to determine the ratio of the two branching ratios
$\BR(\neutralino{1}\rightarrow\muon^{\pm}\W^{\mp})/\BR(\neutralino{1}\rightarrow\tauL^{\pm}\W^{\mp})$, which is related to
the atmospheric neutrino mixing angle $\sin^2 \theta_{23}$.
For an integrated luminosity of $500\ifb$ the total uncertainty on this ratio, including statistical and systematic uncertainties, has been determined to $4\%$.

Finally, we have shown that the precision in measuring the atmospheric neutrino mixing angle is in the same range than measurements from neutrino oscillation experiments, even when
taking parametric uncertainties due to the unknown parts of the SUSY spectrum into account.
Therefore, the International Linear Collider is highly capable to test bRPV SUSY as origin of neutrino masses and mixings.


\section*{Acknowledgement}
We would like to thank the ILC Generators group and the ILD MC production team for providing the SM background samples, as well as Frank G\"ade and the iLCSoft team for 
the great support.
We are particularly grateful to Werner Porod for helpful discussions and for providing the neutrino data fit for the simplified model approach and the input for fig. \ref{fig:parametricuncert}.
We thankfully acknowledge the support by the DFG through the SFB 676 ``Particles, Strings and the Early Universe''.

\appendix
\begin{footnotesize}
\bibliographystyle{apsrev}

\begin{thebibliography}{00}
\bibitem{Wess:1973kz}
  J.~Wess and B.~Zumino,
  Phys.\ Lett.\ B {\bf 49} (1974) 52.

\bibitem{Wess:1974tw}
  J.~Wess and B.~Zumino,
  Nucl.\ Phys.\ B {\bf 70} (1974) 39.
  
 
\bibitem{Haag:1974qh}
  R.~Haag, J.~T.~Lopuszanski and M.~Sohnius,
  Nucl.\ Phys.\ B {\bf 88} (1975) 257.



\bibitem{Romao:1991ex}
  J.~C.~Romao and J.~W.~F.~Valle, 
  Nucl.\ Phys.\ B {\bf 381} (1992) 87.

\bibitem{Mukhopadhyaya:1998xj}
  B.~Mukhopadhyaya, S.~Roy and F.~Vissani,
  Phys.\ Lett.\ B {\bf 443} (1998) 191
  [hep-ph/9808265].
  
  \bibitem{Choi:1999tq}
  S.~Y.~Choi, E.~J.~Chun, S.~K.~Kang and J.~S.~Lee, 
  Phys.\ Rev.\ D {\bf 60} (1999) 075002
  [hep-ph/9903465].

  \bibitem{Romao:2000hv}
    M.~Hirsch, M.~A.~Diaz, W.~Porod, J.~C.~Romao and J.~W.~F.~Valle, 
  Phys.\ Rev.\ D {\bf 62} (2000) 113008
   [Erratum-ibid.\ D {\bf 65} (2002) 119901]
  [hep-ph/0004115].
  
  \bibitem{Porod:2000hv}
  W.~Porod, M.~Hirsch, J.~Romao and J.~W.~F.~Valle, 
  Phys.\ Rev.\ D {\bf 63} (2001) 115004
  [hep-ph/0011248].
  
  \bibitem{Hirsch:2003fe}
  M.~Hirsch and W.~Porod,
  Phys.\ Rev.\ D {\bf 68} (2003) 115007
  [hep-ph/0307364].
  
\bibitem{Behnke:2013lya}
  T.~Behnke, J.~E.~Brau, P.~N.~Burrows, J.~Fuster, M.~Peskin, M.~Stanitzki, Y.~Sugimoto and S.~Yamada {\it et al.},
  ``The International Linear Collider Technical Design Report - Volume 4: Detectors,''
  arXiv:1306.6329 [physics.ins-det].
  
\bibitem{Adolphsen:2013kya}
  C.~Adolphsen, M.~Barone, B.~Barish, K.~Buesser, P.~Burrows, J.~Carwardine, J.~Clark and Hélèn.~M.~Durand {\it et al.},
  ``The International Linear Collider Technical Design Report - Volume 3.II: Accelerator Baseline Design,''
  arXiv:1306.6328 [physics.acc-ph].

  \bibitem{spheno324beta}
  W.~Porod,
  Comput.\ Phys.\ Commun.\  {\bf 153} (2003) 275
  [hep-ph/0301101], 
  W.~Porod and F.~Staub,
  Comput.\ Phys.\ Commun.\  {\bf 183} (2012) 2458
  [arXiv:1104.1573 [hep-ph]],
  and W.~Porod, private communication. 
  
  \bibitem{ATLAS:2012dp}
 [ATLAS Collaboration], 
   \textit{Search for supersymmetry at $sqrt{s} = 7$~TeV~in final states with large jet multiplicity, missing transverse momentum and one isolated lepton with the ATLAS detector},
   ATLAS-CONF-2012-140.
  
\bibitem{LHC:rpvsearches1}
 [ATLAS Collaboration], 
   \textit{Search for strongly produced superpartners in final states with two same sign leptons with the ATLAS detector using 21 fb-1 of proton-proton collisions at sqrt(s)=8 TeV.},
   ATLAS-CONF-2013-007.
\bibitem{LHC:rpvsearches2}
 G.~Aad {\it et al.}  [ATLAS Collaboration], 
   \textit{Search for a heavy narrow resonance decaying to e mu, e tau, or mu tau with the ATLAS detector in sqrt(s) = 7 TeV pp collisions at the LHC},
   arXiv:1212.1272 [hep-ex].
\bibitem{LHC:rpvsearches3}
   G.~Aad {\it et al.}  [ATLAS Collaboration], 
   JHEP {\bf 1212} (2012) 086
   [arXiv:1210.4813 [hep-ex]].
\bibitem{LHC:rpvsearches4}
 [CMS Collaboration], 
   \textit{Search for RPV SUSY resonant second generation slepton production in same-sign dimuon events at sqrt(s) = 7 TeV},
   CMS-PAS-SUS-13-005.
\bibitem{LHC:rpvsearches5}
 [CMS Collaboration], 
   \textit{Search for RPV supersymmetry with three or more leptons and b-tags},
   CMS-PAS-SUS-12-027.
\bibitem{LHC:rpvsearches6}
 [CMS Collaboration], 
   \textit{Search for light stop RPV supersymmetry with three or more leptons and b-tags},
   CMS-PAS-SUS-13-003.
\bibitem{LHC:rpvsearches7}
 [CMS Collaboration], 
   \textit{Search for RPV SUSY in the four-lepton final state},
   CMS-PAS-SUS-13-010.



\bibitem{LHC:directproduction1}
 [CMS Collaboration], 
   \textit{Search for direct EWK production of SUSY particles in multilepton modes with 8TeV data},
   CMS-PAS-SUS-12-022.
\bibitem{LHC:directproduction2}
 [ATLAS Collaboration], 
  \textit{Search for direct production of charginos and neutralinos in events with three leptons and missing transverse momentum in 21$\,$fb$^{-1}$ of pp collisions at $\sqrt{s}=8\,$TeV with the ATLAS detector},
  ATLAS-CONF-2013-035.
\bibitem{LHC:directproduction3}
 [ATLAS Collaboration], 
  \textit{Search for electroweak production of supersymmetric particles in final states with at least two hadronically decaying taus and missing transverse momentum with the ATLAS detector in proton-proton collisions at $\sqrt{s}=8$ TeV},
  ATLAS-CONF-2013-028.

\bibitem{LHC:rpvdirectsearches}
 [ATLAS Collaboration], 
   \textit{Search for supersymmetry in events with four or more leptons in 21$\,$fb$^{-1}$ of pp collisions at $\sqrt{s}=8\,$TeV with the ATLAS detector},
   ATLAS-CONF-2013-036.
   
\bibitem{Baer:2011ec}
  H.~Baer, V.~Barger and P.~Huang,
  JHEP {\bf 1111} (2011) 031
  [arXiv:1107.5581 [hep-ph]].
  
 \bibitem{MoortgatPick:2005cw}
   G.~Moortgat-Pick, T.~Abe, G.~Alexander, B.~Ananthanarayan, A.~A.~Babich, V.~Bharadwaj, D.~Barber and A.~Bartl {\it et al.},
   Phys.\ Rept.\  {\bf 460} (2008) 131
   [hep-ph/0507011].

\bibitem{Bartels:2012ex}
  C.~Bartels, M.~Berggren and J.~List,
  Eur.\ Phys.\ J.\ C {\bf 72} (2012) 2213
  [arXiv:1206.6639 [hep-ex]].

\bibitem{bib:ecalperf}
  C.~Adloff {\it et al.}  [CALICE Collaboration],
  J.\ Phys.\ Conf.\ Ser.\  {\bf 160} (2009) 012065
  [arXiv:0811.2354 [physics.ins-det]].
  
\bibitem{Thomson:2009rp}
  M.~A.~Thomson,
  Nucl.\ Instrum.\ Meth.\ A {\bf 611} (2009) 25
  [arXiv:0907.3577 [physics.ins-det]].
  
  \bibitem{Sarah}
  F.~Staub,
  Comput.\ Phys.\ Commun.\  {\bf 181} (2010) 1077
  [arXiv:0909.2863 [hep-ph]].
  
  \bibitem{Whizard}
  W.~Kilian, T.~Ohl and J.~Reuter, 
  Eur.\ Phys.\ J.\ C {\bf 71} (2011) 1742
  [arXiv:0708.4233 [hep-ph]].\\
  M.~Moretti, T.~Ohl and J.~Reuter, 
  \textit{O'Mega: An Optimizing matrix element generator},
  In *2nd ECFA/DESY Study 1998-2001* 1981-2009
  [hep-ph/0102195].
  
  \bibitem{Pythia}
  T.~Sjostrand, S.~Mrenna and P.~Z.~Skands, 
  JHEP {\bf 0605} (2006) 026
  [hep-ph/0603175].

\bibitem{MoradeFreitas:2002kj}
  P.~Mora de Freitas and H.~Videau,
  ``Detector simulation with MOKKA / GEANT4: Present and future,''
  LC-TOOL-2003-010.
  
  \bibitem{Geant4}
  S.~Agostinelli {\it et al.}  [GEANT4 Collaboration], 
  Nucl.\ Instrum.\ Meth.\ A {\bf 506} (2003) 250.\\
  J.~Allison, K.~Amako, J.~Apostolakis, H.~Araujo, P.~A.~Dubois, M.~Asai, G.~Barrand and R.~Capra {\it et al.}, 
  IEEE Trans.\ Nucl.\ Sci.\  {\bf 53} (2006) 270.
  
\bibitem{Gaede:2006pj}
  F.~G\"ade,
  Nucl.\ Instrum.\ Meth.\ A {\bf 559} (2006) 177.

 \bibitem{bib:MarlinReco}
O. Wendt  \& al.,  Pramana {\bf 69}, (2007) 1109,
 [physics/0702171].
 
\bibitem{Cacciari:2011ma}
  M.~Cacciari, G.~P.~Salam and G.~Soyez,
  Eur.\ Phys.\ J.\ C {\bf 72} (2012) 1896
  [arXiv:1111.6097 [hep-ph]].

\bibitem{hep-ph/0512210}
  M.~Cacciari and G.~P.~Salam,
  Phys.\ Lett.\ B\ {\bf 641} (2006) 57
  [hep-ph/0512210].
 
\bibitem{Beckmann:2010ib}
  M.~Beckmann, B.~List and J.~List,
  Nucl.\ Instrum.\ Meth.\ A {\bf 624} (2010) 184
  [arXiv:1006.0436 [hep-ex]].

 \bibitem{ILDLoI}
  T.~Abe {\it et al.}  [ILD Concept Group - Linear Collider Collaboration], 
  \textit{The International Large Detector: Letter of Intent},
  [arXiv:1006.3396 [hep-ex]].
  

 \bibitem{Porod:2013}
Random scans provided by W.\ Porod, private communication.

   
\bibitem{Fogli:2012ua}
  G.~L.~Fogli, E.~Lisi, A.~Marrone, D.~Montanino, A.~Palazzo and A.~M.~Rotunno,
  Phys.\ Rev.\ D {\bf 86} (2012) 013012
  [arXiv:1205.5254 [hep-ph]].
   
\end{thebibliography}

\end{footnotesize}

\end{document}